\documentclass[11pt]{article}
\usepackage{setspace}
\usepackage{amssymb,amsfonts,amsmath}
\usepackage{cite,enumerate,float,comment}
\usepackage{color}
\usepackage{tikz}
	\usetikzlibrary{backgrounds}
\usepackage{xcolor}
\definecolor{mintcream}{rgb}{0.96, 1.0, 0.98}
\definecolor{champagne}{rgb}{0.97, 0.91, 0.81}
\definecolor{bubblegum}{rgb}{0.99, 0.76, 0.8}
\definecolor{airforceblue}{rgb}{0.36, 0.54, 0.66}
\definecolor{persianblue}{rgb}{0.11, 0.22, 0.73}
\definecolor{zaffre}{rgb}{0.0, 0.08, 0.66}
\usepackage[pdftex,unicode,
colorlinks=true,
linkcolor = zaffre,
citecolor = cyan,
urlcolor = airforceblue]{hyperref}
\usepackage{cleveref}
\usepackage{tcolorbox}

\usetikzlibrary{arrows,snakes,backgrounds}

\def\be{\begin{eqnarray}}
\def\ee{\end{eqnarray}}
\def\nn{\nonumber}

\def\p{\partial}

\definecolor{red}{rgb}{1,0,0}
\definecolor{orange}{rgb}{1,0.5,0}
\definecolor{violet}{rgb}{0.7,0,1}

\usepackage[top = 2 cm, bottom = 2 cm, left = 2.5 cm, right = 1.5 cm]{geometry}

\usepackage[utf8]{inputenc}

\begin{document}

\hfill MIPT/TH-02/24

\hfill ITEP/TH-03/24

\hfill IITP/TH-02/24

\vspace{1cm}
\centerline{\LARGE{
From equations in coordinate space to Picard-Fuchs and back
}}

\bigskip
\vspace{0.5cm}

\centerline{{\bf V.Mishnyakov $^{a,}$\footnote{victor.mishnyakov@su.se}, 	A. Morozov$^{b,c,d,}$\footnote{morozov@itep.ru},	M.Reva$^{b,c,}$\footnote{reva.ma@phystech.edu}}  and {\bf P.Suprun $^{b,}$\footnote{suprun.pa@phystech.edu} } }

\bigskip
\vspace{0.2cm}

\begin{center}
	$^a$ {\small {\it Nordita, KTH Royal Institute of Technology and Stockholm University,}}\\{\small {\it 
Hannes Alfv\'ens v\"ag 12, SE-106 91 Stockholm, Sweden}}
\\
	$^b$ {\small {\it MIPT, Dolgoprudny, 141701, Russia}}\\
        $^c$ {\small {\it NRC ``Kurchatov Institute", 123182, Moscow, Russia}}\\
	$^d$ {\small {\it Institute for Information Transmission Problems, Moscow 127994, Russia}}\\
\end{center} 
\vspace{0.8cm}
       \centerline{\bf \normalsize Abstract}
\vspace{0.8cm}

{
We continue the development of a position space approach to equations for Feynman multi-loop integrals. The key idea of the approach is that unintegrated products of Greens functions in position space are still loop integral in momentum space. The natural place to start are the famous banana diagrams, which we explore in this paper. In position space, these are just products of $n$ propagators. Firstly, we explain that these functions satisfy an equation of order $2^n$. These should be compared with Picard-Fuchs equations derived for the momentum space integral. We find that the Fourier transform of the position space operator contains the Picard-Fuchs one as a rightmost factor. The order of these operators is a special issue, especially since the order in momentum space is governed by degree in $x$ in position space. For the generic mass case this factorization pattern is complicated and it seems like the order of the Fourier transformed position space operators is much bigger than that of the Picard-Fuchs. Furthermore, one may ask what happens if after factorization we take the Picard-Fuchs operators back into position space. We discover that the result is again factorized, with the rightmost factor being the original position space equation. We demonstrate how this works in examples and discuss implications for more sophisticated Feynman integrals.

}

\newpage

\tableofcontents

\bigskip

\section{Introduction}
The study of multi-loop Feynman integrals has been a hot topic in physics lately. This is a very well developed subject therefore we refer to the general reviews \cite{Weinzierl:2022eaz,Vanhove:2018mto,Marcolli:2009zy,Mishnyakov:2022uer}. Differential equations is one of the established methods of studying Feynman integrals. They can be used as a computational tool and an organizational principle.

There are various forms of differential equations, each having its own origin and revealing certain hidden structures. Among those are linear systems obtained by integration-by-part identities \cite{Chetyrkin:1981qh,Argeri:2007up}, GKZ-hypergeometric systems \cite{nasrollahpoursamami2016periods,Vanhove:2018mto,delaCruz:2019skx}, Yangian equations \cite{Loebbert:2022nfu,Kazakov:2023nyu}, Picard-Fuchs equations \cite{Muller-Stach:2012tgj,Lairez:2022zkj,delaCruz:2024xit} and the recently introduced coordinate space equation \cite{Mishnyakov:2023wpd}. All of them are intimately related and have certain advantages and drawbacks. Here we  study the relation between the latter two. Hence, this paper is a continuation of \cite{Mishnyakov:2023wpd} and an extended version of \cite{Mishnyakov:2023sly}. For other usage of the position space approach, see \cite{Cacciatori:2023tzp,Groote:2005ay}.

The overall concept of that we keep mind while developing the position space method is the uplift of differential equations to the level of Ward identities/Schwinger-Dyson equations for correlation functions, and later generic partition functions, similar to Virasoro constraints in the theory of matrix models \cite{morozov1994integrability,Morozov:2005mz}. An extensive discussion of motivations, useful comments, and explanations can be found in \cite{Mishnyakov:2023wpd,Mishnyakov:2022uer}. Here we list the important points  and main theses that are relevant for the present paper.
\begin{itemize}

\item{}
In  {\it coordinate} space we can look at unintegrated correlators. This simplifies the problem, however, doesn't trivialize it: in momentum space these functions are still complicated loop integral. In fact, they are even more general: integration over a vertex in position space amounts to putting the respective outgoing momentum to zero. For \emph{banana}/melon/sunset \footnote{There are several names that have been attached to these graphs depending on the time period and author. We stick to bananas, hoping that this doesn't cause any confusion.} diagrams this is especially relevant, since those naturally do not contain any integration  in coordinate space. In that case, they are just an $n$-fold product of propagators.

\item{}To simplify things it is convenient to consider instead of the true Greens functions the solution of homogeneous equations, i.e. without the delta-function in the r.h.s. In momentum representation this implies the switch from propagators $\frac{1}{p^2-m^2}$ to their imaginary part $\delta(p^2-m^2)$. The total graph is then the \emph{maximal cut} of the original one. From the perspective of differential equations this is in fact not a drastic difference. The differential operator for the cut is the same as for the complete integral, the latter case differs by having an inhomogeneous term. This is true both in position and momentum space \cite{Mishnyakov:2023wpd,Muller-Stach:2012tgj,Vanhove:2018mto}.

\item{} As the cut propagators are defined as solutions of equations of motion one may naturally ask for equations that define products of propagators. The corresponding differential operators are tensor products of equations of motions. These equations  in coordinate representation are straightforward,
still extremely complicated for generic masses.  Since momentum and position space functions are related by Fourier transform, we can apply it to equations as well, thus obtaining a differential equation in momentum space. Note that the idea of Fourier transforming position space equations for Feynman integrals has also emerged in a different setting, mostly related to conformal symmetry \cite{Drummond:2009fd,Coriano:2020ccb,Henn:2023tbo}.

\item{} There is a simple method,  that is nicknamed $\Lambda$-formalism in this text,
which represents these heavily-looking equations as determinants of algorithmically constructed matrices.  After one obtains the position space equation, it can be translated to momentum space by rather simple rules.
The $\Lambda$ formalism is also well suited to Fourier converting the equations into momentum space.

\item{} {\it Momentum} integrals and the maximal cuts satisfy Picard-Fuchs
equations. These can be obtained via the Schwinger parametric representation. This represents the Feynman integral as an integral over a finite number of variables of a singular differential form. The singularity locus defines a  geometry associated to the Feynman integral. From this point of view, the $D=2$ case is special, and it has been determined that the underlying geometry is Calabi-Yau. The Picard-Fuchs equations can then be obtained by methods such as the Griffits-Dwork reduction \cite{Lairez:2022zkj,delaCruz:2024xit}. The appearing equations are quite complicated for $D \neq 2$ and generic masses.

\item{} The main concern of this papers is that the Picard-Fuchs approach can be compared to the Fourier transformed coordinate space. We find that the resulting operators almost exactly match for all masses equal -- the difference being a simple right factor, which is the same for any number of loop. For generic masses the Picard-Fuchs equations appear to be simpler and have much lower order than the Fourier transformed position space equations. Still {\bf the result} is that the equation can be factorized and the Picard-Fuchs part is rightmost irreducible factor:
\begin{tcolorbox}[colback=mintcream]
	\begin{center}
	Fourier transform of coordinate-space eq-n = Complicated $\times$  Picard-Fuchs
	\end{center}
\end{tcolorbox}
or in our notations
\begin{equation}
\mathcal{\hat{E}}_x^{(n)} \, \longrightarrow \, \hat{E}^{(n)}_t  =  \hat{\bf{B}} \cdot \widehat{PF} \,.
\end{equation}

\item It appears that for generic masses the degree of $\hat{E}_t$ in  $\frac{\p}{\p t}$ is much bigger then that of $\widehat{PF}$. The conjectured degree of $\widehat{PF}$ grows exponentially with $n$, and we observe that  $\hat{\bf{B}}$ seems to grow at least quadratically  with the number of loops.
\end{itemize}

\bigskip

In what follows, we will illustrate these theses, dividing them into different blocks.
Each block contains somewhat borrowing calculations, and ends with the summary of
the relevant morals.

\bigskip

In sec.\ref{sec:coord} we review the coordinate-space equations for banana FD from \cite{Mishnyakov:2023wpd}
and the $\Lambda$-formalism to formulate them conceptually and deduce practically.
We explain, why the product of Green functions
\be
B_n(x) := \prod_{i=1}^n G_{m_i}(x)
\label{coordprod}
\ee
satisfies differential equation of order $2^n$, and describe technical ways to derive
them in explicit form. The resulting equation is:
\begin{equation}
    \hat{\mathcal{E}}^{(n)}_x B_n(x)=0
\end{equation}
While conceptually trivial, the equations are pretty complicated beyond one loop,
and no generic expression is yet available.
Considerable simplifications arise when some masses are equal,
and the order drops to $n$ when they are all the same.
\\

In sec.\ref{sec:PF} we go through the derivation and properties of Picard-Fuchs operators in momentum space.
The resulting equation will be denoted
\be
\widehat{PF} \cdot  B_n(t) = 0.
\ee

In sec.\ref{sec:xtop} we explain a way to convert the equations to momentum space,
i.e. to substitute $x^\mu$ by $\p_{p^\mu}$ and $p^\mu \cong \p_{x^\mu}$ and $B_n(x)$ by $B_n(t)$
with $t:= p^\mu p_\mu$. It is straightforward in $\Lambda$-formalism.
Since derivatives and powers are interchanged by Fourier transform,
the order in momentum space is defined by the power in $x$ of the original equation
(not its degree in $\p_x$).
The resulting equations will be denoted as:
\begin{equation}
	\hat{E}^{(n)}_t \cdot B_n(t)=0
\end{equation}

In sec.\ref{sec:factorization} we address the main issue of this paper --
the relation between the two operators $\hat E_t$ and $\widehat{PF}$. We observe the factorization property:
\be
\hat{E}^{(n)}_t  =  \hat{\bf{B}} \cdot \widehat{PF}.
\label{factoriz}
\ee
There are no {\it well established} way to study algebraic factorization of sophisticated differential operators,
thus getting (\ref{factoriz}) is quite an interesting problem.
In addition, $\hat{\bf{B}}$ need not be (and is not) {\it polynomial} in $t$ --
its coefficients are rather rational functions of $t$.
Thus, a better version  of (\ref{factoriz}) is
\be
Q(t) \cdot \hat{E}^{(n)}_t = \hat {\cal B} \cdot \widehat{PF}
\label{factorizP}
\ee
where $Q(t)$ is some polynomial and $\hat {\cal B} = Q(t) \cdot  \hat{\bf{B}} $  has polynomial coefficients.
\\

We also explore the inverse relation with the Fourier transform of the Picard-Fuchs operator into position space. It appears that the resulting operator, which we denote $\widehat{PF}_x$ also factorizes, with the rightmost factor being the initial position space equations. 
\\

Finally, in sec.\ref{beyond} we touch the possibility to generalize
the whole story to less trivial diagrams, than the banana family.
The product  formulas like (\ref{coordprod}) remain true whenever
we keep the positions of all vertices fixed (not integrated over),
but the algorithm to find differential equations in coordinate space
is not yet found in full generality. At the same time, PF equations in momentum space are always available --
and can be associated with various families of curves and higher-dimensional
complex Calabi-Yau manifolds --
this part of the story is well publicized in the literature \cite{Bonisch:2021yfw,Bourjaily:2019hmc,Bourjaily:2018yfy}. Promotion to the level of factorization relations like
(\ref{factoriz}) and the search of an adequate version of (\ref{factorizP})
are the challenging tasks for the future work.
\\

We end with a short conclusion in sec. \ref{sec:conclusion} where we discuss the main challenges and future directions.

\section{Coordinate-space equation
	\label{sec:coord} }

\subsection{$\Lambda$-formalism}\label{lform}

In \cite{Mishnyakov:2023wpd} a method of computing coordinate--space differential equations was proposed. Here we present a modification of it, which seems to be more relevant for the purposes of this article. But first, let us review the idea in simple terms. The basic observation is that in coordinate space the banana graph functions is a product of propagators:
\begin{equation}
	B_n(x) = \prod_{i=1}^{n} G_{m_i}(x)
\end{equation}
where each propagator satisfies the equation of motion:
\begin{equation}
	\left(\Box + m^2 \right) G_m(x)= \delta(x-y)
\end{equation}
As announced in the introduction we consider instead a simplified problem with propagator substituted with solutions to the homogeneous equations of motion, which in momentum space corresponds to taking the maximal cut. With a slight abuse of notation we denote it by the same letter and only consider the maximal cuts further on:
\begin{equation}\label{eq:KG}
	\left(\Box + m^2 \right) G_m(x)=0
\end{equation}
If we consider only invariant solutions then  it is just a second order equation that has a two dimensional space of solutions. Knowing that, we can construct the tensor product of two such differential operators with different masses, such that the kernel of the new operator is the four possible products of solutions of the initial equation. This can be easily demonstrated in $D=1$, where \eqref{eq:KG} is just:
\begin{equation}
	\left( \dfrac{\partial^2}{\partial x^2}+m^2\right)G_m(x)=0 \,  \Rightarrow \, G_m(x) \sim e^{\pm i m x}
\end{equation}
There are four 2-banana functions in this case:
\begin{equation}
	B_2(x)\sim e^{\pm i m_1 x \pm i m_2 x}
\end{equation}
The equation in question is then:
\begin{equation}
	\left( \dfrac{\partial^2}{\partial x^2} + (m_1+m_2)^2\right)	\left( \dfrac{\partial^2}{\partial x^2} + (m_1-m_2)^2\right) B_2(x)= 0
\end{equation}
This can be easily generalized to any $n$ if we keep $D=1$.
For generic $D$ the equation is more complicated. If we denote $x=\sqrt{x^\mu x_\mu}$, then:
\begin{equation}
	\left(\Box+m^2\right)G_m(x)=\left(\dfrac{\partial^2}{\partial x^2 } + \dfrac{D-1}{x} \dfrac{\partial}{\partial x}+ m^2\right) G_m(x)=0
\end{equation}
The solutions to these equations are given by Bessel functions, however, this knowledge does help much in deriving the equation for their product. We won't need the explicit form of the solution below. Instead, we provide a procedure to derive the equations. Some examples of the derivation can be found in \cite{Mishnyakov:2023wpd}. Here we proceed with the general case immediately. The main difference here is the choice of generators for the ring of differential operators: while  in \cite{Mishnyakov:2023wpd} the resulting equation was expressed using three operators: $x^2, \Lambda = x^\mu \dfrac{\partial}{\partial x^\mu}$ and $\Box$, here we stick to using powers of $x^2$ and $\Lambda$  instead. The two representations are related by the identity
\be
\Lambda^2 = x^2\, \hat\Box - (D-2)\, \Lambda
\ee
which follows from the way the operators act on the function of modulus $x$,
\be
\p_{\mu} f\left(x=\sqrt{x^\mu x_\mu}\right) = \frac{x^\mu}{x} \cdot \dfrac{\partial f(x)}{\partial x}  \ \ \Longrightarrow \ \
\begin{split}
	&\Lambda  f = x \partial_x f \\
	&\Lambda^2 f = x^2 \partial^2_x f +x  \partial_x f \\
	&\hat\Box f = \partial^2_x f + \frac{D-1}{x}  \partial_x f
\end{split}
\ \ \Longrightarrow \ \  x^2 \Box F =  \Lambda^2 F + (D-2)\Lambda F
\ee
and allows one to express powers of $\Lambda$ in terms of powers $\hat\Box$ and no more than one $\Lambda$, or vice versa. While the equations that make use of the $\Box$ operator as well as $\Lambda$ from \cite{Mishnyakov:2023wpd} are more compact,
an expression in terms of $\Lambda$ only  is more suitable for general reasoning, as we will now demonstrate.
\\

Let us rewrite the equation of motion one last time using only $\Lambda$ operator:
\begin{equation}
	(\Box + m^2) G_m (x) = \frac{1}{x^2}(\Lambda^2 + (D-2)\Lambda + x^2 m^2)G_m (x) = 0.
\end{equation}
We see that in order to have a polynomial operator we have to multiply by $x^2$ from the left. We will see some consequence of this later. Now we present a procedure to obtain a differential equation of the banana function.
\\

Firstly, we notice that one can reduce any power of the dilatation operator acting on the propagator to first derivatives and multiplication by $x^2$ using the equations of motion. Namely:
\begin{equation}\label{eq:LambdaOnG}
	\Lambda^k G_m =  \sum_{i=0}^{k} a_{k,i} x^{2i} \Lambda G_m + \sum_{i=1}^k b_{k,i} x^{2i}  G_m
\end{equation}
There does not seem to be any simple expression for these coefficients, however they can be defined recursively. Next, let us introduce a notation for a product of $\Lambda$-derivatives of the propagators:
\begin{equation}
	I_{\vec{k}}=I_{\left(k_1,\ldots k_n \right)}:= \prod_{i=1}^{n} \Lambda^{k_i} G_{m_i}
\end{equation}
where $k_i$ can be zero. In these notations the initial banana function is just:
\begin{equation}
	B_n(x)=I_{\vec{0}}:= I_{\left(0,\ldots \right)}
\end{equation}
Thus, applying the dilatation operator to the product of propagators we get:
\begin{equation}\label{eq:LambdaOnI1}
	\Lambda^n I_{\vec{0}} = \sum_{\vec k : \sum k_i=n} \binom{n}{k_1,\ldots k_n} I_{\vec{k}}
\end{equation}
Using the expansion \eqref{eq:LambdaOnG} for each of the $I_{\vec{k}}$ functions we obtain:
\begin{equation}\label{eq:LambdaOnI2}
	\Lambda^k I_{\vec{0}}=  \sum_{\vec{\epsilon}\,:\,\epsilon_i={0,1} } \!\! C_{k}^{\vec{\epsilon}} I_{\vec{\epsilon}}
\end{equation}
where the coefficients $C_{k}^{\vec{\epsilon}}$ are made up of the $a_{k,i},b_{k,i}$  of \eqref{eq:LambdaOnG} and the multinomial coefficients in \eqref{eq:LambdaOnI1}.
\\

The key idea is to notice that these relations for $k=1,\ldots 2^n$  imply that the homogeneous linear system with matrix elements $ \left(C_{k}^{\vec{\epsilon}} | \Lambda^{k} I_{\vec{0}}  \right)$ has a nontrivial solution given by the $2^{n}+1$--ple $\left( \left\{I_{\vec{\epsilon}} \right\}_{\epsilon_i={0,1}} , -1 \right)$
This means that the determinant of this matrix should vanish which implies a differential equation on $I_{\vec{0}}=B_n$:
\begin{equation}\label{eq:DeterminantEq}
	\mathcal{E}^{(n)}_x I_{\vec{0}}:=\det_{\substack{k = 0,\ldots, 2^n \\ \vec{\epsilon} \in \mathbb{Z}_2^n }} \left( C_{k}^{\vec{\epsilon}} | \Lambda^{k} I_{\vec{0}}  \right)  = 0
\end{equation}

The coefficients $C_{k}^{\vec{\epsilon}}$ are in principle computed via the recursion for and $a_{k,i},b_{k,i}$, however, there is no explicit formula for them in terms of the masses and dimensions. Clearly, the final determinant is a more complicated object. However, formula \eqref{eq:DeterminantEq} establishes the existence and the form of the position space equations for generic $n$-banana functions with distinct masses. We will later demonstrate how one can obtain explicit equations from this determinant and even deduce their general form for the massless case.
\\

One property of the resulting equation that we are interested in is the degree of its coefficients in $x$. The equation has the form:
\begin{equation}
	\sum_{i=0}^{2n} \alpha_i(x) \Lambda^i = 0
\end{equation}
where $\alpha_i(x)$ are clearly just the minors that appear in the last column expansion of the determinant in \eqref{eq:DeterminantEq}. The degrees in $x^2$ of subsequent coefficients are related \begin{equation}
	\deg_{x^2} \left( \alpha_{i+1} \right) =\deg_{x^2} \left( \alpha_i \right)  - 2(i \mod 2) 
\end{equation}
However, we are yet unable to provide a general formula for this degree. For the lowest $n$ we have:
\begin{center}
        \begin{tabular}{c|c|c|c|c|c}
     $n$ &  $1$ & $2$ & $3$ & $4$ & $5$\\
    $\deg_{x^2} \left( \alpha_{2n}(x) \right) $& $0$& $4$& $34$& $196$ & $964$ 
    \end{tabular}
\end{center}
This growth look somewhat like $2^{2 n}-2^n$.
\\

To illustrate the construction, consider the example of $n=2$ generic mass banana diagram. In this case, the basic function is $I_{0,0}=B_2 = G_{m_1} G_{m_2} := G_1 G_2 $ and there are three ways to apply the dilatation operators:
\[ I_{1,0} = (\Lambda G_1) G_2;\ I_{0,1} = G_1 (\Lambda G_2);\ I_{1,1} = (\Lambda G_1) (\Lambda G_2). \] To get an equation we need to apply derivatives to the banana function four times. By iterated application of $\Lambda$ operator and using the relations \eqref{eq:LambdaOnG} we obtain:
\begin{equation}\label{eq:n2example}
	\begin{split}
		& \phantom{{}^0}\Lambda I_{0,0} = I_{1,0} + I_{0,1} \\
		& \Lambda^2 I_{0,0} = 2 I_{1,1} -(D-2)I_{1,0} - (D-2) I_{0,1} - (m_1^2 + m_2^2 )x^2 I_{0,0}  \\
		& \Lambda^3 I_{0,0} = -6 (D-2)I_{1,1} - (x^2 m_1^2  + 3 x^2 m_2^2 - (D-2)^2)I_{1,0} -
		\\
		& \phantom{\Lambda^4 I_{0,0}=} - (3 x^2 m_1^2 + m_2^2 x^2 -(D-2)^2)I_{0,1} + (D-4) (m_1^2 + m_2^2) x^2 I_{0,0} \\
		& \Lambda^4 I_{0,0} = (-4 (m_1^2+m_2^2) x^2+14 (D-2)^2) I_{1,1} +(2 x^2 ((D-4)m_1^2 + (5D-14)m_2^2) -(D-2)^3)I_{1,0} +\\
		& \phantom{\Lambda^4 I_{0,0}=}+(2 x^2 ((5D-14) m_1^2+(D-4) m_2^2)-(D-2)^3) I_{0,1} +\\
		& \phantom{\Lambda^4 I_{0,0}=} (-(m_1^2+m_2^2)(D^2 -6 D +12)+ 6
		m_2^2 m_1^2 x^2+m_1^4 x^2+m_2^4 x^2) I_{0,0} \\
	\end{split}
\end{equation}
Where, for instance, the second line was obtained as follows:
\begin{equation}
	\Lambda^2 I_{0,0} = I_{2,0} + 2 I_{1,1}+ I_{0,2}
\end{equation}
and
\begin{equation}
	\begin{split}
		&I_{2,0} = x^2 (D-2) I_{1,0} + m_1^2 I_{0,0}
		\\
		&I_{0,2} = x^2 (D-2) I_{1,0} + m_2^2 I_{0,0}
	\end{split}
\end{equation}
The matrix $\left(C_{k}^{\vec{\epsilon}} | \Lambda^{k} I_{\vec{0}}  \right)$ is then explicitly given by:
{\small
\[ \begin{pmatrix}
	1 & 0 & 0 & 0 & I_{0,0} \\
	0 & 1 & 1 & 0 & \Lambda I_{0,0} \\
	- (m_1^2 + m_2^2 )x^2 & -(D-2) & -(D-2) & 2 & \Lambda^2 I_{0,0} \\
	(D-4)(m_1^2 + m_2^2)x^2 & - (m_1^2 x^2 + 3 m_2^2 x^2 -(D-2)^2) & -(3 m_1^2 x^2 + m_2^2 x^2 - (D-2)^2) & -6(D-2) & \Lambda^3 I_{0,0} \\
	C_4^{00} & C_4^{10} & C_4^{01} & C_4^{11} & \Lambda^4 I_{0,0} \\
\end{pmatrix}\]
}
where
\begin{equation}
	\begin{split}
		C_4^{00} &=  x^2 (-(m_1^2+m_2^2)(D^2 -6 D +12)+ 6
		m_2^2 m_1^2 x^2+m_1^4 x^2+m_2^4 x^2),	 \\
		C_4^{10} &= (2 x^2 ((D-4)m_1^2 + (5D-14)m_2^2) -(D-2)^3),  \\
		C_4^{01} &=(2 x^2 ((5D-14) m_1^2+(D-4) m_2^2)-(D-2)^3),  \\
		C_4^{11} &= -4 (m_1^2+m_2^2) x^2+14 (D-2)^2
	\end{split}
\end{equation}
The equations \eqref{eq:n2example} imply that the matrix has a kernel of the form $(I_{0,0}, I_{1,0}, I_{0,1}, I_{1,1},-1)$.
Therefore, its determinant vanishes, which produces an equation:
\begin{equation} \label{1loopcoord}
	\begin{aligned}
		\mathcal{E}^{(2)}_x I_{\vec{0}} =& 4 x^2 (m_1^2 - m_2^2) \Big\{ \Lambda^4 + 2 (2 D - 5) \Lambda^3  +
		\Big(2 x^2 (m_1^2 + m_2^2)+(D-2)(5D-16)\Big) \Lambda^2  +
		\\
		&+ 2 \Big((m_1^2+m_2^2) x^2 (2D-3)+(D-4) (D-2)^2 \Big) \Lambda  + \\
        &+ x^2 \Big((m_1^2 -m_2^2)^2 x^2 + 2 (D-1)(D-2) (m_1^2 +m_2^2)\Big) \Big\} I_{0,0} = 0
	\end{aligned}
\end{equation}

As we see, the equation is of order $4$ in $x$ derivatives just as expected. Another important feature is the factorization of $x^2$, which is similar to the same phenomenon in the equation of motion. Moreover, if we recall that when acting on invariant functions $\Lambda=x\partial_x$ we can actually factorize another power of $x$ from the left of the differential operator.

\subsection{Primary examples}
Here we list examples of position space differential operators for various cases.
\subsubsection{Different masses}
	
\paragraph{One loop, $n=2$.} 	As we just obtained in (\ref{1loopcoord}), the operator, annihilating the product $G_{m_1}\cdot G_{m_2}$, is
	\be
	\mathcal{E}^{(2)}_x  \sim	\Lambda^4  + 2 (2 D - 5) \Lambda^3   +
	\Big(2 x^2 (m_1^2 + m_2^2)+(D-2)(5D-16)\Big) \Lambda^2  +
	\nn
	\ee
	\begin{equation}
		+ 2 \Big((m_1^2+m_2^2) x^2 (2D-3)+(D-4) (D-2)^2 \Big) \Lambda  +
		x^2 \Big((m_1^2 -m_2^2)^2 x^2 + 2 (D-1)(D-2) (m_1^2 +m_2^2)\Big)
		\label{diff_mass_coord_1_loop}
	\end{equation}

\paragraph{Two loops, $n=3$.}
	The triple product $G_{m_1}\cdot G_{m_2}\cdot G_{m_3}$
	is annihilated by a similar, but far more sophisticated operator. To write it out, introduce a notation:
	\begin{equation}
		M_{i,j,k}= m_{1}^i m_2^j m_3^k + \text{symm}
	\end{equation}
	for the symmetric sums of masses. The 3 mass differential operator then reads:
	\begin{equation}
		\begin{aligned}
			& \mathcal{E}^{(3)}_x \sim \Big(-8 (D-3) (D-1) x^2 M_{0,0,2}+16 x^4 (M_{0,0,4}-2 M_{0,2,2})-
			\\
		& - 3 (D-7) (D-3) (D-1) (D+3) \Big)\Lambda^8 +\\
			& + 4\Big( 16 (3 D-10) x^4 (M_{0,0,4}-2 M_{0,2,2})-4 (D-3) (D-1) (6 D-19) x^2 M_{0,0,2}-\\
			&-9 (D-7) (D-3)^2 (D-1) (D+3) \Big)  \Lambda^7+ \sum_{i=0}^6 q_i(x,\vec{m},D) \Lambda^i
		\end{aligned}
		\label{diff_mass_coord_2_loops}
	\end{equation}
	The coefficients get quite large and complicated already in this case. For example:
	{\small
	\begin{equation*}
		\begin{split}
			q_3(x,&\vec{m},D) =2 \Big(96 (D-2) x^8 (3
			M_{0,0,8}+2 (-2 M_{0,2,6}+M_{0,4,4}-6 M_{2,2,4}))\\
			&\qquad -8 x^6 (-18 (D (78 D^2-639 D+1682)-1421) M_{2,2,2}+(D (D (258 D-2053)+5278)-4383) M_{0,0,6}+ \\
			&\qquad \qquad (D ((1781-210 D) D-4814)+4143) M_{0,2,4})+ \\
			&\qquad 2 x^4 ((D (D (D (D (1389 D-21967)+139922)-441506)+683521)-413887) M_{0,0,4}+\\
			&\qquad \qquad 2 (D (D (D ((11339-465 D)
			D-91642)+335050)-572021)+370267) M_{0,2,2}) \\
			&\qquad  -4 (D-3) (D-1) (3 D (D (3 D (D (16 D-333)+2696)-30050)+51379)-98342) x^2 M_{0,0,2} \\
			&\qquad  -9 (-7 + D) (-3 + D)^2 (-2 + D) (-1 + D) (3 + D) (-1664 +
			D (1500 + D (-428 + 39 D))) \Big)
		\end{split}
	\end{equation*}
}

Deriving an equation for higher $n$ is equally straightforward, but the answers become way to long to fit in a paper.

\subsubsection{Equal masses}
We continue with examples for equal masses, where we know the equations to be of smaller order. This happens, because in this case the matrices degenerate. These result reproduce the equations from \cite{Mishnyakov:2023wpd}, now in a unified framework.
\begin{itemize}
	
	\item{$n=2$.}
	For equal masses the above matrix becomes
	\begin{equation}
		\left(
		\begin{array}{ccccc}
			1 & 0 & 0 & 0 &  I_{0,0} \\
			0 & 1 & 1 & 0 &  \Lambda^1 I_{0,0} \\
			-2 m^2 x^2 & 2-D & 2-D & 2 &  \Lambda^2 I_{0,0} \\
			2 (D-4) m^2 x^2 & (D-2)^2-4 m^2 x^2 & (D-2)^2-4 m^2 x^2 & -6 (D-2) &  \Lambda^3 I_{0,0}  \\
			C_4^{00} & C_4^{10} & C_4^{01} & C_4^{11} & \Lambda^4 I_{0,0}  \\
		\end{array}
		\right)
	\end{equation}
	While it does not seem to simplify a lot its determinant is now identically zero. This is one of the manifestation of the fact that the true equation for the equal mass case is of lower order. Indeed, now:
	\begin{equation}
		I_{1,0} = I_{0,1}
	\end{equation}
	and $C_4^{10}=C_4^{01}$. Deleting the last row and say the third column we get:
	\begin{equation}
		\left(
		\begin{array}{cccc}
			1 & 0 & 0 &  I_{0,0} \\
			0 & 1 & 0 & \Lambda^1 I_{0,0} \\
			-2 m^2 x^2 & 2-D & 2 & \Lambda^2 I_{0,0} \\
			2 (D-4) m^2 x^2 & (D-2)^2-4 m^2 x^2 & -6 (D-2) & \Lambda^3 I_{0,0} \\
		\end{array}
		\right)
	\end{equation}
	
			Taking the determinant we obtain the differential equation:
			\begin{equation}
				\mathcal{E}^{(2)}_x \sim	\left(\Lambda ^3+3 (D-2) \Lambda ^2+  \left(2 (D-2)^2+4 m^2 x^2\right) \Lambda +4 (D-1) m^2 x^2 \right)
				\label{eq_mass_coord_1_loop}
			\end{equation}
			Finally, one can also consider the limit of vanishing masses in when the equations becomes:
			\begin{equation}
				\Lambda  (\Lambda +D -2) (\Lambda+2D -4) I_{0,0}= 0
			\end{equation}
			We will analyze this limit in greater detail below.

			\item{$n=3$}
			
			The matrix has the form {\small
			\[ \hspace{-1.1cm}  \begin{pmatrix}
				1 & 0 & 0 & 0 & G_3 \\
				0 & 3 & 0 & 0 & \Lambda G_3 \\
				- 3 m^2 x^2 & 6 - 3 D & 6 & 0 & \Lambda^2 G_3 \\
				3 (D-4) m^2 x^2 & - (7 m^2 x^2 -(D-2)^2) & -18 (D-2) & 6 & \Lambda^3 G_3 \\
				3 m^2 x^2 (7 m^2 x^2 - D(D-6) -12) & 6 (11D-32) m^2 x^2 - 3(D-2)^2  & -60 m^2 x^2 + 42(D-2)^2 & - 36 (D-2) & \Lambda^4 G_3 \\
			\end{pmatrix}\]
		}
			The resulting determinant gives
			\begin{equation}
				\begin{aligned}
					&\mathcal{E}^{(3)}_x \sim 108 \Lambda^4 + 648 (D-2) \Lambda^3 + (1188 (D-2)^2 + 1080 m^2 x^2) \Lambda^2 +\\
					&+ 216 (3(D-2)^3 + 5(3D-4) m^2 x^2 )\Lambda +  324 m^2 x^2 (3 m^2 x^2 + 2 D (3D-7) + 8)
				\end{aligned}
				\label{eq_mass_coord_2_loops}
			\end{equation}
			
		\end{itemize}
		
		We don't list the equations further as a lot of examples were given in \cite{Mishnyakov:2023wpd} in the $\Box, \Lambda$ notation.
		
		\subsubsection{Vanishing masses
		}
		Taking the limit of vanishing masses greatly simplifies the calculations. It is to be considered a ''solvable'' example: we can completely derive the form of the general position space operator starting from the equations of motion.  In this case the cut propagator satisfies:
		\be
		\Big(\Lambda^2 + (D-2)\Lambda\Big)G = 0
		\ee
		which makes applying powers of $\Lambda$ trivial:
		\begin{equation}
			\Lambda^{n}G = (2-D)^{n-1} \Lambda G \, , n>1
		\end{equation}
		By Leibniz rule we now simply have:
		\begin{equation}
			\begin{split}
				\Lambda^k G^n = &\sum_{\lambda_1,\lambda_2,\ldots \lambda_n \geq 0} \binom{k}{\lambda_1 \ \lambda_2 \ \ldots \ \lambda_n} \prod_{i=1}^{n} \Lambda^{\lambda_i} G =\\
				& = \sum_{p=1}^{k} \sum_{\lambda_1 ,\ldots, \lambda_p \geq 1}  \binom{k}{\lambda_1 \ \lambda_2 \ \ldots \ \lambda_p } \prod_{i=1}^{n}(D-2)^{\lambda_i} G^{n-p} (\Lambda G)^{p}
			\end{split}
		\end{equation}
		Here $\binom{k}{\lambda_1 \ \lambda_2 \ \ldots \ \lambda_n}$ are the multinomial coefficients
		\begin{equation}
			\binom{k}{\lambda_1 \ \lambda_2 \ \ldots \ \lambda_n} = \dfrac{k!}{\lambda_1! \cdot \ldots \cdot \lambda_n!}
		\end{equation}
		Using:
		\begin{equation}
			\sum_{\lambda_1,\ldots \lambda_n \geq 1} \binom{r}{\lambda_1 \ \lambda_2 \ \ldots \ \lambda_n} = n! S_2(r,n) \, \ r>n
		\end{equation}
		where $S_2(r,n)$ is the Stirling number of the second kind we obtain:
		\begin{equation}
			\Lambda^k G_n = \sum_{i=1}^{\min\left(k,n \right)} (2-D)^{n-i+1} i!S(k,i) \binom{n}{n-i} G^{n-i}(\Lambda G)^{i}
		\end{equation}
		One has the following matrix:
		\begin{equation}
			\left( M_n \right)_{i,j}=  (2-D)^{n-i+1} i!S(j,i) \binom{n}{n-i}
		\end{equation}
		Determinant is given by:
		\begin{equation}
			\mathcal{E}_x^{n} = \det M_n = \left(\prod_{i=0}^{n} \dfrac{n!}{(n-i)!} \right) \det M_n^{(\text{red})}
		\end{equation}
		where the reduced matrix is given by:
		\begin{equation}
			M_n^{(\text{red})}=\left\{\left( S_2(i,k) (2-D)^{n-i+1}| \Lambda^{i} \right) \right\}_{\substack{i=1,\ldots, n+1 \\
					k=1,\ldots ,  n
			} }
		\end{equation}
		This determinant makes sense since the coefficients are now independent of $x$, hence no issues arrive with non-commutativity of operators. One can easily show that this determinant has zeros,  if we formally substitute $\Lambda=p (D-2)$ for $p=0,\ldots n$. This is done by proving that at these values the columns are linearly dependent via an identity:
		\begin{equation}
			\sum_{i=0}^p S_2(k,i) \dfrac{p!}{(p-i)!} = p^k
		\end{equation}
		On the other hand, the determinant is a polynomial of degree $n+1$ in $\Lambda$. Therefore, we have obtained all zeros of the polynomial in $\Lambda$, and we can immediately write:
		\begin{equation}
			\hat{\mathcal{E}}_x^{(n)}  =  \left(\prod_{i=0}^{n} \dfrac{n!}{(n-i)!} \right) \left(\prod_{i=0}^{n}(\Lambda+i(D-2))\right)
		\end{equation}
		As before vanishing of the determinant gives a differential equation:
		\be
		\hat{\mathcal{E}}_x^{(n)}   B_n(x) =0
		\ee
		Let us again illustrate this with two simple examples:
		\begin{itemize}
			\item At $n=2$ the derivatives are given by
			\begin{equation}
				\begin{split}
					\Lambda  G^2 &= 2G\Lambda G \nn \\
					\Lambda^2 G^2 &= 2G\Lambda^2 G + 2(\Lambda G)^2 = -2(D-2)G\Lambda G + 2(\Lambda G)^2 \nn \\
					\Lambda^3 G^2 &= -2(D-2)G\Lambda^2 G - 2(D-2)(\Lambda G)^2 +4\Lambda G \Lambda^2 G =
					2(D-2)^2 G \Lambda G - 6(D-2)(\Lambda G)^2
				\end{split}
			\end{equation}
			And the matrix:
			\begin{equation}
				M_2=\left(\begin{array}{ccc}
					2 & 0 & \Lambda Z_2 \\
					-2(D-2) & 2 & \Lambda^2 Z_2 \\
					2(D-2)^2 & -6(D-2) & \Lambda^3 Z_2 \\
				\end{array}\right)
			\end{equation}
			Hence the determinant gives
			\be
			\mathcal{E}_x^{(2)} = -4\Lambda\Big(\Lambda^2 + 3(D-2)\Lambda +2(D-2)^2\Big)
			= -4\Lambda \Big(\Lambda+(D-2)\Big) \Big(\Lambda+ 2(D-2)\Big)
			\label{vanishing_1_loop_coord}
			\ee
			
			\item At $n=3$ we similarly have:
			\begin{equation}
				\begin{split}
					&	\Lambda G^3 = 3G^2\Lambda G, \nn \\
					&\Lambda^2 G^3 = - 3(D-2)G^2\Lambda G + 6G(\Lambda G)^2, \nn \\
					&\Lambda^3 G^3 = +3(D-2)^2 G^2\Lambda G - (6+12)(D-2)G(\Lambda G)^2 + 6(\Lambda G)^3, \nn \\
					&\Lambda^4 G^3 = -3(D-2)^3 G^2\Lambda G + (6+2\cdot 18 )(D-2)^2G(\Lambda G)^2 -(18+18)(D-2)(\Lambda G)^3
				\end{split}
			\end{equation}
			and
			\begin{equation}
				\begin{gathered}
					M_3 =  \left(\begin{array}{cccc}
						3 & 0 & 0 & \Lambda Z_3 \\
						-3(D-2) & 6 & 0 & \Lambda^2 Z_3 \\
						3(D-2)^2 & -18(D-2) & 6 & \Lambda^3 Z_3 \\
						-3(D-2)^3 & 42(D-2)^2 & -36(D-2) & \Lambda^4 Z_3 \\
					\end{array}\right)
					\\
					\\
					\mathcal{E}_x^{(3)} = 108\cdot \Lambda\Big(\Lambda^3
					+ 6 (D-2)\Lambda^2 + 11 (D-2)^2\Lambda +6 (D-2)^3 \Big) = \nn \\
					= \dfrac{3!3!3!3!}{0!1!2!3!}\cdot \Lambda \Big(\Lambda+(D-2)\Big) \Big(\Lambda+ 2(D-2)\Big) \Big(\Lambda+3(D-2)\Big)
				\end{gathered}
			\end{equation}
		\end{itemize}

\section{Momentum space Picard-Fuchs operators
\label{sec:PF}}

It is quite well known by now, that momentum space banana integrals satisfy peculiar Picard-Fuchs equations. These follow from the Schwinger parametric presentation. These equations were studied extensively in \cite{Muller-Stach:2012tgj,Lairez:2022zkj,delaCruz:2024xit}. Here we review the whole procedure in technical terms and provide a variety of examples, that should be compared later with Fourier transforms of coordinate space operators.
\\

Before going on to describe the equations, we want to briefly discuss the issue of ambiguities in the choices of the momentum space propagator. This issue will also be mentioned in section \ref{sec:numberofsolutions}. To simplify the discussion we deal with $D=1$. The problem in question is the equation:
\begin{equation}
    (p^2-m^2)G(p)=0
\end{equation}
A generic solution can be expressed as:
\begin{equation}
    G(p) = a \delta(p-m) + b\delta(p+m)
\end{equation}
For the equation for the true propagator the same issue will be present as there will be an ambiguity in the choice of the solution to the homogeneous equation. The issue is well known and the proper field theory in four dimensional Minkowski space corresponds to the difference between advanced, retarded or causal Greens functions. Here we simply want to highlight that below we will be discussing the loop integrals where $G(p)$ is the cut causal propagator, in $D=1$ meaning that we choose:
\begin{equation}
    G(p) = \dfrac{\delta(p-m)}{2m}+\dfrac{\delta(p+m)}{2m} = \delta(p^2-m^2)
\end{equation}
Which is a specific combination of basis solutions. These choices can be otherwise encoded in a choice of function $f(p)$ that would multiply the solution:
\begin{equation}
    G(p)=f(p)\delta(p^2-m^2)
\end{equation}
By choosing $f(p)$ such that its support will only intersect one of the points in the delta function, we can choose between different solutions. To see the effect of this on the loop integrals, consider a one loop example:
\begin{equation}
    B_2(x) = \int dp e^{ipx} \int dq f_1(p-q)f_2(q)  \delta(q^2-m^2)\delta((p-q)^2-m^2) = \int dp e^{ipx} f(p) B_n(p) 
\end{equation}
Where $B_n(p)=\int dq \delta(q^2-m^2)\delta((p-q)^2-m^2)$.
\\

We summarize be stressing that in momentum space that different solutions of the momentum space equations correspond to the choice of weights for each support of the delta-functions. These can be encoded as a contour choice or a choice of function that specifies these weights. This should be taken into account when discussing the order of the differential operators in this section, and further for comparison with position space.

\subsection{Picard-Fuchs equations for Feynman diagrams}
In what follows we neglect the above-mentioned ambiguities and
consider the equations in the momentum space,
derived {\it not} by a Fourier transform from the configuration space,
but directly from the momentum integral.
We remind that our consideration in this paper is restricted to
{\it homogeneous} Green functions, which in momentum space are delta-functions
rather than more familiar Feynman propagators.
In that case, they are equations for the multiple integral, depending on a parameter $t=p^2$,
and have their own order in $t$ and $\p_t$ -- not directly related to the order of
the equation in coordinate space. In this section we are interested in deriving a homogeneous linear differential equation on the banana integral in momentum space $B_n(t)$. The main idea to represent it as an integral of some differential form over a closed domain, then to construct an operator sending the form to a closed one. It implies that the same operator nullifies $B_n(t)$. Proceeding with the manipulations in detail, we get:
\begin{equation}
	\begin{aligned}
		&B_n(t=p^2)= \int \limits_{\mathbb{R}^{D^n}}  \delta\left(\sum_{i=1}^n\vec k_i - \vec p\right)
\prod_{i=1}^n \delta(k_i^2-m_i)^2 d\vec k_i=  \\
&= \prod_{i=1}^n \int\limits_{\mathbb{R}^D} d\vec k_i\int\limits_{\mathbb{R}^{D+n}}  e^{i\sum_{i=1}^n a_i(k_i^2-m_i^2)
			+ i\vec\beta\left(\sum_{i=1}^n\vec k_i - \vec p\right)}
		d^D\beta \prod_{i=1}^n d a_i=\\
		&=\int\limits_{\mathbb{R}^{D+n}} \prod_{i=1}^n da_i   e^{-\sum_{i}a_i m_i^2-\vec{\beta}\vec{p}}
\left(\prod_{i=1}^n \int\limits_{\mathbb{R}^{n}}
e^{i(a_i\vec{k_i}^2+\vec{\beta}\vec{k_i})} d\vec{k_i}\right)  d^D\beta   =\\
		&=\int\limits_{\mathbb{R}^{D+n}}
\left(e^{\sum_{i=1}^n a_i^{-1} (\frac{\vec{\beta}}{2})^2 -\vec{\beta}\vec{p}} \, d^D\beta\right)
\prod_{i=1}^n a_i^{-\frac{D}{2}}e^{-a_i m_i^2} d a_i
 =\\
		&=\int\limits_{\mathbb{R}^n} \prod_{i=1}^n d a_i
\left(\prod_{i=1}^n a_i\sum_{i=1}^n a_i^{-1}\right)^{-\frac{D}{2}}
e^{\frac{\vec{p}^2}{\sum a_i^{-1}}-\sum_{i=1}^n a_i m_i^2}= \backslash a_{i}=\lambda \alpha_i \backslash=\\
		&=\int\limits_{\mathbb{R}^n}\prod_{i=1}^{n}d \alpha_i \delta\left(1-\sum_{i=1}^{n} \alpha_i\right)
\left(\prod_{i=1}^n \alpha_i\sum_{i=1}^n \alpha_i^{-1}\right)^{-\frac{D}{2}}
\ \int\limits_{0}^{+\infty} d\lambda\ \lambda^{(n-1)\frac{2-D}{2}}e^{\lambda(\frac{\vec{p}^2}{\sum_{i=1}^n \alpha_i^{-1}}-\sum_{i=1}^n \alpha_i m_i^2)}=\\
		&=\int\limits_{\mathbb{R}^n}\prod_{i=1}^n d \alpha_i \delta\left(1-\sum_{i=1}^n \alpha_i\right) \frac{\left(\prod_{i=1}^{n}\alpha_i\sum_{i=1}^{n}\alpha_i^{-1}\right)^{\frac{n}{2}(2-D)}}{\left(\left(\prod_{i=1}^{n} \alpha_i\sum_{i=1}^{n} \alpha_i^{-1}\right)\left(\sum_{i=1}^{n}m_i^2 \alpha_i\right)-t\prod_{i=1}^{n} \alpha_i\right)^{n-\frac{(n-1)D}{2}}}=\\
		&=\int\limits_{\mathbb{R}^n}\prod_{i=1}^n d \alpha_i
\delta\left(1-\sum_{i=1}^n \alpha_i\right)\frac{U^{\frac{n}{2}(2-D)}}{F^{1-\frac{(n-1)(D-2)}{2}}}
	\end{aligned}
	\label{Shwinger_rep}
\end{equation}
where $U=\prod_{i=1}^{n}\alpha_i\sum_{i=1}^{n}\alpha_i^{-1}$ and $F=\left(\prod_{i=1}^{n} \alpha_i\sum_{i=1}^{n} \alpha_i^{-1}\right)\left(\sum_{i=1}^{n}m_i^2 \alpha_i\right)-t\prod_{i=1}^{n} \alpha_i$
are called the first and the second Symanzik polynomials respectively.
The main advantage of this representation is the homogeneous integrand
(up to the $\delta$-function),
i.e. does not change under the transformation
$(\alpha_1,\dots,\alpha_n)\rightarrow (\lambda \alpha_1,\dots,\lambda\alpha_n)$.
That is the reason the integral can be written as a projective one:
\begin{equation}
	B_n=\int\limits_{\Gamma^n}\frac{U^{\frac{n}{2}(2-D)}}{F^{1-\frac{(n-1)(D-2)}{2}}}\,\omega
	\label{ProjI_rep}
\end{equation}
where $\Gamma_n=\left\{ (\alpha_1,\dots,\alpha_n)\in \mathbb{CP}^{n-1}\vert (\alpha_1,\dots,\alpha_n)\in \mathbb{R}^n \right\}$ and
$\omega=\sum_{i=1}^n (-1)^{i+1}\alpha_i d\alpha_1\wedge\dots\wedge\widehat{d\alpha}_i\wedge\dots\wedge d\alpha_n$
is the standard projective measure.
Since we were discussing the maximal cut we obtain a closed contour $\Gamma_n$  and it allows to construct Picard-Fuchs operator for this integral
\begin{equation}
	\left(\sum_{i=1}^{k}a_i(t)\frac{\partial^i}{\partial t^i}\right)\cdot B_n(t)=\widehat{PF} \cdot B_n(t)=0 \hspace{5mm}\Longleftrightarrow \hspace{5mm} \widehat{PF} \cdot \left(\frac{U^{\frac{n}{2}(2-D)}}{F^{1-\frac{(n-1)(D-2)}{2}}}\omega \right)= d\beta
	\label{PF_proj_def}
\end{equation}
It is more convenient to reformulate this equation in terms of complex integral as the underlying differential form will become simpler.
\begin{equation}
	B_n(t)=\frac{1}{2\pi i} \int\limits_{\widetilde{\Gamma}_n} \frac{U^{\frac{n}{2}(2-D)}}{F^{1-\frac{(n-1)(D-2)}{2}}} \prod_{i=1}^{n}d \widetilde{\alpha}_i
	\label{ComplexI_rep}
\end{equation}
where $\widetilde{\Gamma}_n$ is the fibration over $\Gamma_n$ with fibre $S^1$: $\widetilde{\Gamma}_n=\left\{ (\widetilde{\alpha}_1,\dots,\widetilde{\alpha}_n) \in\mathbb{C}^n\ \vert\ \widetilde{\alpha}_i =\alpha_i e^{i\phi},\ \alpha_i \in \Gamma_n,\ \phi \in S^1 \right\}$. To check \eqref{ComplexI_rep}
it is convenient to use the identity
\be
\prod_{i=1}^nd \widetilde{\alpha}_i =\frac{1}{n}\left(\frac{d\widetilde{\alpha}_1}{\widetilde{\alpha}_1}+\dots + \frac{d\widetilde{\alpha}_n}{\widetilde{\alpha}_n}\right) \wedge
\left(\sum_{i=1}^n (-1)^{i+1} \widetilde{\alpha_i} d\widetilde{\alpha}_1\wedge\dots\wedge\widehat{d\widetilde{\alpha}}_i\wedge\dots
\wedge d\widetilde{\alpha}_n\right)
\ee
Integrating over the fibre results in \eqref{ProjI_rep}:
\begin{equation}
	B_n(t)= \frac{1}{2\pi} \int\limits_{\widetilde{\Gamma}_n} \frac{U^{\frac{n}{2}(2-D)}}{F^{1-\frac{(n-1)(D-2)}{2}}} d\phi\wedge
 \left(\sum_{i=1}^n (-1)^{i+1}\alpha_i d\alpha_1\wedge\dots\wedge\widehat{d\alpha}_i\wedge\dots \wedge d\alpha_n\right)
\end{equation}
which is an obvious identity. Thus \eqref{PF_proj_def} transforms to
\begin{equation}
	\widehat{PF} \cdot B_n(t)=0 \hspace{5mm}\Longleftrightarrow \hspace{5mm} \widehat{PF} \cdot \left(\frac{U^{\frac{n}{2}(2-D)}}{F^{1-\frac{(n-1)(D-2)}{2}}} \right)= \sum_{i=1}^{n}\frac{\partial g_i(\alpha)}{\partial \alpha_i}
	\label{PF_complex_def}
\end{equation}
We will further omit the difference between $\widetilde{\alpha}_i$ and $\alpha_i$, and $\widetilde{\Gamma}_n$ and $\Gamma_n$.

\subsection{Primary examples}
The simplest example is $1$-loop banana integral $(n=2)$, which can be expressed as
\begin{equation}
	B_2(t)=\int\limits_{\Gamma_n}\frac{\left(\alpha_1+\alpha_2\right)^{2-D}}{\left((\alpha_1+\alpha_2)(m_1^2\alpha_1+m^2_2\alpha_2)-t\alpha_1\alpha_2\right)^{2-\frac{D}{2}}}d\alpha_1d\alpha_2
\end{equation}
To derive Picard-Fuchs operator it is important to find suitable ansatz for $g_i$ in r.h.s of \eqref{PF_complex_def}. For arbitrary $D$ we can consider the ansatz
\begin{equation}
	g_i=f_i \frac{U^a}{F^b}
	\label{g_i_init}
\end{equation}
where $f_i$ is some polynomial in $\alpha$, $U=\alpha_1+\alpha_2$ and $F=(\alpha_1+\alpha_2)(m_1^2\alpha_1+m^2_2\alpha_2)-t\alpha_1\alpha_2$. Then, after substituting $\widehat{PF}=\sum_{i=0}^k a_i\frac{\p}{\p t}$ and \eqref{g_i_init} in \eqref{PF_complex_def} we get
\begin{equation}
\begin{aligned}
    \left((2-\frac{D}{2})\dots(1+k-\frac{D}{2})(\alpha_1\alpha_2)^k a_k+\dots+a_0  F^k\right) & \frac{U^{2-D}}{F^{2+k-\frac{D}{2}}}= \\
    &=\sum_{i=1}^2\left(\frac{\partial f_i}{\partial \alpha_i}U F+a f_i \frac{\partial U}{\partial \alpha_i}F-b f_i U \frac{\partial F}{\partial \alpha_i} \right) \frac{U^{a-1}}{F^{b+1}}
	\label{ansatz_cond_one_loop}
\end{aligned}
\end{equation}
Transforming this equation we arrive at
\begin{equation}
	\left(A-B\frac{U^{a+D-3}}{F^{b-k+\frac{D}{2}-1}}\right)\frac{U^{2-D}}{F^{2+k-\frac{D}{2}}}=0
\end{equation}
where $A=\left((2-\frac{D}{2})\dots(1+k-\frac{D}{2})(\alpha_1\alpha_2)^k a_k+\dots+a_0  F^k\right)$ and $B=\sum_{i=1}^2\left(\frac{\partial f_i}{\partial \alpha_i}U F+a f_i \frac{\partial U}{\partial \alpha_i}F-b f_i U \frac{\partial F}{\partial \alpha_i} \right)$. As both $U$ and $F$ are non-zero we have
\begin{equation}
	A-B\frac{U^{a+D-3}}{F^{b-k+\frac{D}{2}-1}}=0
\end{equation}
This implies, that $B\frac{U^{a-D-3}}{F^{b-k+\frac{D}{2}-1}}$ is a polynomial which is satisfied in two cases
\begin{equation}
	\left[
	\begin{array}{ll}
		B=k\, F^{b-k+\frac{D}{2}-1}, & \text{where}\ k\ \text{is a polynomial} \\
		b-k+\frac{D}{2}-1=0 &
	\end{array}
	\right .
\end{equation}
We want $B$ to be a polynomial not divisible by $F$ and $U$ which implies $b=k-\frac{D}{2}+1$. As the result we are left with
\begin{equation}
	A-B\, U^{a+D-3}=0 \hspace{5mm} \rightarrow \hspace{5mm} \frac{A}{U^{a-D-3}}-B=0
\end{equation}
Again we have two possibilities
\begin{equation}
	\left[
	\begin{array}{ll}
		A=k\, U^{a-D-3}, & \text{where}\ k\ \text{is a polynomial} \\
		a+D-3=0 &
	\end{array}
	\right .
\end{equation}
We, again, want $A$ to be a polynomial not divisible by $F$ and $U$ which implies $a=3-D$. However, this ansatz will work only in $D>2$ case. For $D=2$ there is no $U$ in l.h.s. of \eqref{ansatz_cond_one_loop} which is the reason we put $a=0$ in \eqref{g_i_init} to find a non-trivial solution. Below we will consider the case $D=2$ separately.
\subsubsection{$D>2$}
After setting  $a=3-D$ and $b=1+k-\frac{D}{2}$ we arrive at the equation,
\begin{equation}
\begin{aligned}
	\left(2-\frac{D}{2}\right)\dots\left(1+k-\frac{D}{2}\right)(\alpha_1\alpha_2)^k & a_k  +\dots+a_0  F^k= \\
 & =\sum_{i=1}^2\left(\frac{\partial f_i}{\partial \alpha_i}U F+\left(3-D\right) f_i \frac{\partial U}{\partial \alpha_i}F-\left(1+k-\frac{D}{2}\right) f_i U \frac{\partial F}{\partial \alpha_i} \right)
	\label{f_i_eq_d_one_loop}
\end{aligned}
\end{equation}
From this equation we see that $f_i$ can be found as a homogeneous polynomial in $\alpha_1$ and $\alpha_2$ of degree $2(k-1)$. Thus $f_i=\sum\limits_{j=0}^{2(k-1)}f_{i,j}\alpha_1^i\alpha_2^{2(k-1)-i}$ has $2(k-1)+1$ undetermined coefficients $f_{i,j}$ which can be defined by transforming \eqref{f_i_eq_d_one_loop} as
\begin{equation}
	\sum_{i=0}^{2(k-1)}c_i\alpha_1^i\alpha_2^{2(k-1)-i}=0\hspace{5mm}\rightarrow\hspace{5mm} \begin{cases}
		c_0=0\\
		\dots\dots\\
		c_{2(k-1)}=0
	\end{cases}
	\label{lin_sys_one_loop_D}
\end{equation}
The lowest $k$ when this system has non-trivial solution is $k=1$, so the resulting Picard-Fuchs operator has order $1$ and is presented below. For convenience we also present the r.h.s. of \eqref{PF_complex_def}.
\begin{equation}
	\begin{aligned}
		&\left\{2t(t-(m_1-m_2)^2)(t-(m_1+m_2)^2) \frac{\partial}{\partial t} + \left( (D-2)(m_1^2-m_2^2)^2-2(m_1^2+m_2^2)t+ (4-D)t^2 \right)\right\} \frac{U^{2-D}}
		{F^{2-\frac{D}{2}}}= \\
		&=	2\left(m_2^2(t+m_1^2-m_2^2)\frac{\partial}{\partial \alpha_1}+m_1^2(t+m_2^2-m_1^2)\frac{\partial}{\partial \alpha_2}\right)\frac{U^{3-D}}
		{F^{2-\frac{D}{2}}}
	\end{aligned}
	\label{PF_one_loop_diff_mass_arbitrary_D}
\end{equation}
In the case of equal masses operator simplifies, and we arrive at
\begin{equation}
	\left\{t(t-4m^2)\frac{\p}{\p t} -(D-4)t -4m^2\right\}\frac{U^{2-D}}{F^{2-\frac{D}{2}}}= m^2\left(\frac{\p}{\p \alpha_1}+\frac{\p}{\p \alpha_2}\right)\frac{U^{3-D}}{F^{2-\frac{D}{2}}}
    \label{PF1loopequalmasses}
\end{equation}
Until now we worked with coordinates in $\mathbb{C}^2$. However, as explained in the previous subsection, it is not the only way. That is why we present this example reformulated in terms of Schwinger representation \eqref{Shwinger_rep}.\\
\be
\left(\frac{\p}{\p t}
- \frac{(D-4)t^2 + 2(m_1^2+m_2^2)t - (D-2) (m_1^2-m_2^2)^2}{2 t \big(t-(m_1+m_2)^2\big) \big(t-(m_1-m_2)^2\big)}\right)
\frac{1}{\left(a(1-a)p^2-am_1^2-(1-a)m_2^2\right)^{2-\frac{D}{2}}} =  \nn \\
= \frac{1}{t\big(t-(m_1+m_2)^2\big)\big(t-(m_1-m_2)^2\big)}
\frac{\p}{\p a} \left\{\frac{ua+v}{\left(a(1-a)t-am_1^2-(1-a)m_2^2\right)^{2-\frac{D}{2}}}\right\}
\label{PF1loopdiffmasses}
\ee
where $u=-(m_1^2+m_2^2)t +(m_1^2-m_2^2)^2$ and $v= m_2^2(t+m_1^2-m_2^2)$.
From the l.h.s. of above identity we get the PF equation for $B_2(t)$:
\be
\left\{2t\big(t-(m_1-m_2)^2\big)\big(t-(m_1+m_2)^2\big)\frac{\p}{\p t} -
(D-4)t^2 -2(m_1^2+m_2^2)t +(D-2)(m_1^2-m_2^2)^2\right\} B_2(t) = 0
\ee
For equal masses $m_1=m_2=m$ (\ref{PF1loopdiffmasses}) reduces to just
\be
\left(\frac{\p}{\p t} - \frac{(D-4)t+4m^2}{2t(t-4m^2 )}\right)
\frac{1}{\left(\alpha _1\alpha _2t-m^2\right)^{2-\frac{D}{2}}} \
+ \left. \frac{m^2}{t(t-4m^2)}
\left(\frac{\p}{\p \alpha _1}-\frac{\p}{\p \alpha _2}\right)
\frac{\alpha _1-\alpha _2}{\left(\alpha _1\alpha _2 t-m^2\right)^{2-\frac{D}{2}}}
\right|_{\alpha _2=1-\alpha _1}\!\!\!\!\!   = 0
\ee

\subsubsection{$D=2$}
In this case the former ansatz \eqref{ansatz_cond_one_loop} will change as $U$ vanishes in l.h.s.
\begin{equation}
	\left(k!(\alpha_1\alpha_2)^k a_k+\dots+a_0  F^k\right)\frac{1}{F^{1+k}}= \sum_{i=1}^2\left(\frac{\partial f_i}{\partial \alpha_i} F-b f_i  \frac{\partial F}{\partial \alpha_i} \right) \frac{1}{F^{b+1}}
	\label{ansatz_cond_one_loop_d_2}
\end{equation}
After putting $b=k$ we arrive at
\begin{equation}
	k!(\alpha_1\alpha_2)^k a_k+\dots+a_0  F^k=\sum_{i=1}^2\left(\frac{\partial f_i}{\partial \alpha_i} F-k f_i  \frac{\partial F}{\partial \alpha_i} \right)
	\label{f_i_eq_d=2_one_loop}
\end{equation}
Repeating the procedure from the previous subsection we find that $f$ is a homogeneous polynomial of degree $2k-1$. The lowest $k$ for which this equation can be solved is $k=1$ and the resulting operator reads
\begin{equation}
	\left((t-(m_1-m_2)^2)(t-(m_1+m_2)^2) \frac{\partial}{\partial t}+t-m_1^2-m_2^2\right) \frac{1}{F}=\left(m_2^2\frac{\p}{\p \alpha_1}+m_1^2\frac{\p}{\p \alpha_2}\right)\frac{1}{F}
\end{equation}
In the case of equal masses this equation simplifies to
\begin{equation}
	\left((t(t-4m^2)\frac{\p}{\p t}+t-2m^2\right)\frac{1}{F}=m^2\left(\frac{\p}{\p \alpha_1}+ \frac{\p}{\p \alpha_2}\right)\frac{1}{F}
	\end{equation}
In Schwinger representation \eqref{Shwinger_rep} the differential equation can be written in the form:
\be
\left(\frac{\p}{\p t} + \frac{t-m_1^2-m_2^2}{\big(t-(m_1+m_2)^2\big)\big(t-(m_1-m_2)^2\big)}\right)
\frac{1}{a(1-a)t-am_1^2-(1-a)m_2^2} =  \nn \\
= \frac{1}{t\big(p^2-(m_1+m_2)^2\big)\big(t-(m_1-m_2)^2\big)}
\frac{\p}{\p a} \left\{\frac{ua+v}{a(1-a)t-am_1^2-(1-a)m_2^2}\right\}
\label{PF1loop_Schwinger}
\ee
with $u=-(m_1^2+m_2^2)t +(m_1^2-m_2^2)^2$ and $v= m_2^2(t+m_1^2-m_2^2)$.\\
In equal mass case the equation will be
\begin{equation}
\left(\frac{\p}{\p t} + \frac{t-2m^2}{t\left(t-4m^2\right)}\right)
\frac{1}{a(1-a)t-m^2} = \frac{1}{t\left(t-4m^2\right)}
\frac{\p}{\p a} \left\{\frac{m^2(1-2a)}{a(1-a)t-m^2}\right\}
\label{PF1loop_equal_Schwinger}
\end{equation}

\subsection{Calabi-Yau case}
As we have shown in previous section the case $D=2$ is quite special. The reason behind it is the appearance of a (singular) Calabi-Yau variety.
\begin{equation}
	B_n(t)\Bigg|_{D=2}=\Omega=\int_{\Gamma_n} \frac{\omega}{F} \hspace{5mm} \rightarrow \hspace{5mm} X:\ F=0
\end{equation}
$F$ is homogeneous and the number of variables $(\alpha_i)$ and degree of $F$ is $n$, so the canonical bundle is trivial. This variety $X$ is singular for $n\geq 4$, so in general it is a one-parametric family with respect to parameter $t$ of (singular) Calabi-Yau varieties. This fact allows getting more properties of Picard-Fuchs operator of the underlying Feynman integral. Further in this section we will work only in $D=2$.\\

Denote a generic operator by $P=\sum\limits_{i=0}^n a_i\frac{\partial^i}{\partial t^i}$ and its formal adjoint by $P^*=\sum\limits_{i=0}^n (-1)^i \frac{\partial^i}{\partial t^i}a_i$, then the following relation holds:
\begin{equation}
	\label{adjoint_eq}
	Pf(t)=(-1)^{\text{deg} P}f(t)P^{*}
\end{equation}
for some function $f(t)$. To understand the source of this equation we give an example of an order 2 operator. We have the Kronecker pairing
\begin{equation}
	H_{t,k}\times H^{k}_t\rightarrow \mathbb{C}[t]
\end{equation}
Applying Poincare duality we arrive at intersection form
\begin{equation}
	\begin{aligned}
		&H_t^{2n-k}\times H^k_t\rightarrow \mathbb{C}[t]\\
		&(\phi,\theta)=\int \phi\wedge\theta
	\end{aligned}
\end{equation}
In our case form $\Omega$ lies in $H^{n,0}$, that's why we are interested in map
\begin{equation}
	H_t^{n}\times H_t^{n,0}\rightarrow\mathbb{C}[t]
\end{equation}
There are multiple choices for $\phi \in H_t^n$, however if $Pr_{H_t^{n,0}}\phi=0$ then $(\phi,\theta)=0$ for $\theta \in H_t^{n,0}$. Moreover, $X$ is of Calabi-Yau type so $\dim H_t^{n,0}=1$. To make computation more concrete we consider the 2-nd order operator $P$ and the following expression:
\begin{equation}
	(P f(t)\Omega,\Omega)= a_2\left(\frac{\partial^2}{\partial t^2}(f\Omega),\Omega \right)+a_1 \left(\frac{\partial}{\partial_t}(f\Omega),\Omega \right)+a_0(\Omega,\Omega)
\end{equation}
Continuously applying Leibniz rule and permuting the form we arrive at:
\begin{equation}
	\label{form_eq_duality}
	(P f(t)\Omega,\Omega)= (-1)^{deg P} (fP^*\Omega,\Omega)+(f(a_1-a_2')+f'a_2)'(\Omega,\Omega)
\end{equation}
Hence to get an equation on operators it's crucial to choose $f$ satisfying $(f(a_1-a_2')+f'a_2)'=0$ and applying the fact that $\dim H^{n,0}_t=1$ we get the desired equation. \\
In general case this equation will transform to
\begin{equation}
	f(2 a_{n-1}-n a_n')+n a_n f'=0
\end{equation}
 This equation always have a solution, so the following operator identity is satisfied:
\begin{equation}
	\label{eq_duality}
	P f=(-1)^{\text{deg}P}f P^*
\end{equation}
We have conducted these calculations for all the operators present in this paper and found out that operators in $D> 2$ do not satisfy this property but have similar structure. In particular we find that for the equal mass banana operators one has (see sec.\ref{sec:factorization} and  \cite{Mishnyakov:2023sly})
\begin{equation}\label{eq:coefrelforequallmassDnot2}
	a_{n-2}= \dfrac{(n-1)}{2} \cdot \left( \dfrac{n(D-2)}{2t} a_{n-1} -(D-3) \dfrac{\partial a_{n-1} }{\partial t} \right)
\end{equation}

\subsection{Equal masses - up to five loops}
Previously we stated, that the ansatz \eqref{g_i_init} with $a\neq 0$ will always suffice for $D> 2$. However, it turns out, that for equal mass case this ansatz will work in $D=2$ case too, which significantly simplifies the computations. In this section we present the differential operators, sometimes with r.h.s., for equal masses banana integrals in arbitrary dimension $D$ and list some of their properties.
\paragraph{2 loops }
The Symanzik polynomials for the 2 loop banana integral are
\be
U = \alpha_2\alpha_3+\alpha_1\alpha_3+\alpha_1\alpha_2
\ee
\be
F=(\alpha_2\alpha_3+\alpha_1\alpha_3+\alpha_1\alpha_2)(\alpha_1+\alpha_2+\alpha_3)m^2 - t \cdot\alpha_1\alpha_2\alpha_3
\ee
Reproducing the computations in (\ref{f_i_eq_d_one_loop}-\ref{lin_sys_one_loop_D}) we arrive at big linear system, which rank is much lower than the number of equations. That is the reason for big gauge freedom in the chose of r.h.s. We choose the most symmetric gauge which satisfies the following property
\begin{equation}
	f_1(\alpha_1,\alpha_2,\alpha_3)=f_1(\alpha_1,\alpha_3,\alpha_2)=f_2(\alpha_2,\alpha_1,\alpha_3)=f_2(\alpha_2,\alpha_3,\alpha_1)=f_3(\alpha_3,\alpha_1,\alpha_2)=f_3(\alpha_3,\alpha_2,\alpha_1)
	\label{symmetric_f_i}
\end{equation}
This condition will still leave some gauge freedom, which we do not know how to fix in an invariant way. Finally, the operator in a certain gauge reads
{\footnotesize
\begin{equation}
    \begin{split}
    	&\Bigg\{2 t\,(t-m^2)(t-9m^2)\frac{\p^2}{\p t^2} + \Big(-3(D-4)t^2+10(D-6)m^2t+9     D m^4\Big)\frac{\p}{\p t}+
     \\
     &\hspace{7cm}
	+(D-3)\Big((D-4)t+(D+4)m^2\Big) \Bigg\}\frac{U^{\frac{3}{2}(2-D)}}{F^{3-D}}
	=  \\
	&= 2(D-3)m^2 \Bigg\{ \frac{\p}{\p \alpha_1}\frac{ \Big( \alpha_1^2(t-              m^2)-2m^2\alpha_2\alpha_3  \Big)
		U^{\frac{3}{2}(2-D)+1} }{F^{4-D}}
	+\frac{\p}{\p \alpha_2}  \Big( \alpha_2^2(t-m^2)-2m^2\alpha_1\alpha_3         \Big)  +    \\
        &\hspace{10cm}+\frac{\p}{\p \alpha_3} \Big( \alpha_3^2(t-m^2)-2m^2\alpha_1\alpha_2  \Big) \Bigg\} \dfrac{U^{\frac{3}{2}(2-D)+1} }{F^{4-D}}
	\label{pf_2_loops_equal}        
    \end{split}
\end{equation}
}

\paragraph{3 loops}
The Symanzik polynomials for 3 loop banana integral are
\be
U = \alpha_2\alpha_3\alpha_4+\alpha_1\alpha_3\alpha_4+ \alpha_1\alpha_2\alpha_4+\alpha_1\alpha_2\alpha_3
\ee
\be
F=(\alpha_2\alpha_3\alpha_4+\alpha_1\alpha_3\alpha_4+ \alpha_1\alpha_2\alpha_4+\alpha_1\alpha_2\alpha_3)
(\alpha_1+\alpha_2+\alpha_3+\alpha_4)m^2 - t \cdot \alpha_1\alpha_2\alpha_3\alpha_4 
\ee
We repeat the computations as in the previous subsection and get the operator:
\be
\left\{4t^2(t-4m^2)(t-16m^2)\frac{\p^3}{\p t^3} + 12 t \Big(-(D - 4)t^2 + 10 (D-5) m^2 t + 64 m^4\Big)\frac{\p^2}{\p t^2}
+\right. \nn \\
+ \Big((D - 4) (11 D - 36)t^2 - 4  (7 D^2- 88 D + 216)m^2t - 
 64 D (D-4) m^4\Big)\frac{\p}{\p t}
+\nn \\
\left.-(D-3) (3 D - 8) \Big( (D-4)t + 2 (D+2) m^2)\Big) \right\}\frac{U^{2(2-D)}}{F^{4-\frac{3}{2}D}}
=  \sum_{i=1}^4 \frac{\p}{\p\alpha_i}  \frac{f_i\, U^{5-2D} }{F^{6-\frac{3}{2}D}}
\label{eq_3_loops_pf}
\ee
where the gauge is the same as before. The first term at the r.h.s. contains
\begin{equation}
    \begin{aligned}
        f_1=&(3D-8)m^2\Big\{ 6(3D-10)m^2\alpha_2^2\alpha_3^2\alpha_4^2 + \\
        &+2\left((7D-24)t+(7D-18)m^2\right) m^2 \alpha_1\alpha_2\alpha_3\alpha_4 (\alpha_2\alpha_3 + \alpha_2\alpha_4 + \alpha_3\alpha_4) +\\
        &+\left((D-4)t+2(D+2) m^2\right)m^2 \alpha_1\alpha_2\alpha_3\alpha_4 (\alpha_2^2+\alpha_3^2+\alpha_4^2)+\\
        &+\left(2(D-3)t^2-(17D-44) m^2 t-(6D-52)m^4\right) \alpha_1^2\alpha_2\alpha_3\alpha_4(\alpha_2+\alpha_3+\alpha_4)+\\
        &+\left((7D-24) t-4 (D-6) m^2 \right)m^2\alpha_1^2 (\alpha_2^2\alpha_3^2+\alpha_2^2\alpha_4^2+\alpha_3^2\alpha_4^2)+\\
        &+\left((D-4)t+2 (D+2)m^2\right)m^2 \alpha_1^2(\alpha_2^3\alpha_3+\alpha_2\alpha_3^3 +\alpha_2^3\alpha_4+\alpha_2\alpha_4^3 +\alpha_3^3\alpha_4+\alpha_3\alpha_4^3)\Big\}
    \end{aligned}
\end{equation}
which is a function of the $6$-th order in $\alpha$, symmetric under the permutations of $\alpha_{2,3,4}$
(since all masses are the same).
Other $f_i$ are obtained by permutations of $\alpha$'s.
\paragraph{4 loops}
The Symanzik polynomials for 4 loop banana integral are
\begin{equation}
U = \alpha_2\alpha_3\alpha_4\alpha_5+\alpha_1\alpha_3\alpha_4\alpha_5+ \alpha_1\alpha_2\alpha_4\alpha_5+\alpha_1\alpha_2\alpha_3\alpha_5+ \alpha_1\alpha_2\alpha_3\alpha_4
\end{equation}
\begin{equation}
F=U(\alpha_1+\alpha_2+\alpha_3+\alpha_4+\alpha_5)m^2- t\cdot \alpha_1\alpha_2\alpha_3\alpha_4\alpha_5 
\end{equation}
In this subsection and further on we will omit r.h.s of \eqref{PF_complex_def} as it becomes very complicated
\begin{equation}
	\begin{aligned}
		\Big\{&-16 t^2 (t-25 m^2 ) ( t-9 m^2) (t-m^2)\frac{\partial^4}{\partial t^4}+ \\
		& +16 t \Big( 5 (D-4) t^3- 35 (3D-14 ) m^2 t^2+ 259 (D-8) m^4 t+225 (D+2) m^6\Big)\frac{\partial^3}{\partial t^3}+\\
        &+4 \Big(- 5 (D-4) (7D-24) t^3 + 7 (D-6) (49 D-172) m^2 t^2 +\\
        &+ 3 ( D (552 + 121 D)-4040) m^4 t + 225 D (D + 2) m^6 \Big)\frac{\partial^2}{\partial t^2}+\\
		&+4 (D-3) \Big(5 (D-4) (5D-16) t^2- 14 (D-16) (3D-10) m^2 t-3 D (69 D-328) m^4  \Big)\frac{\partial}{\partial t}-\\
        &-4 (D-3) (2D-5) (3D-8) ((D-4) t + (3D+4) m^2)\Big\}B_5=0
	\end{aligned}
	\label{eq_4_loops_pf}
\end{equation}

\paragraph{5 loops}
\begin{equation}
	\begin{aligned}
		\Big\{&8 t^3 (t-36 m^2) (t-16 m^2) (t-4 m^2)\frac{\partial^5}{\partial t^5}-\\
        &-20 t^2 \Big(3 (D-4) t^3 - 56 (2D-9 ) m^2 t^2 + 784 (D-6) m^4 t + 6912 m^6\Big)\frac{\partial^4}{\partial t^4}+\\
        &+2 t \Big( 5 (D-4) (17 D-60) t^3
		-56 (624 + D (33 D-292)) m^2 t^2+\\
		& +48 (3861 + 2 D (31 D-614)) m^4 t + 6912 (D-6) (D+2) m^6\Big)\frac{\partial^3}{\partial t^3}+\\
        &+\Big( - 15 (D-4) (3 D-10) (5 D-16) t^3 +	56 (2 D-7) (432 + D (19 D-196)) m^2 t^2  +\\
        &+ 16 (22104 + D (-7735 + 4 D (76 D-177))) m^4 t + 2304 D (D+2) (2 D-7) m^6\Big)\frac{\partial^2}{\partial t^2}-\\
        &-( D-3) \Big(- (D-4) (1200 +	D (137 D-812)) t^2+ 8 (-4800 + D (3284 + D (13 D-597))) m^2 t +\\
		& +16 D (1391 + 8 D (11 D-93)) m^4 \Big)\frac{\partial}{\partial t}-\\
		&-(D-3) (2 D-5) (3 D-8) ( 5 D-12) \Big((D-4) t + 4 (D+1) m^2\Big)\Big\}B_6=0
	\end{aligned}
	\label{eq_5_loops_pf}
\end{equation}

\paragraph{Higher loops.}
For high loop number finding the PF equation directly becomes technically difficult. We provide the 6 and 7 loop data in Appendix  \ref{AppendixA1}
One can notice that certain terms of these equations have a simple structure. We list these patterns, that were first observed in \cite{Lairez:2022zkj}. See also sec.\ref{sec:xtop} and \cite{Mishnyakov:2023wpd,Mishnyakov:2023sly} for the derivation of these from coordinate space.
\begin{itemize}
	\item the order of equation is $n-1$ and it does not depend on $D$
	\item the degree in $t$ of the leading coefficient equals $n$
	\item the leading coefficient has the form $a_{n-1}=t^{\lceil\frac{n}{2}\rceil}\prod\limits_{i=1}^{\lceil\frac{n+1}{2}\rceil}(t-(2i-1)^2 m^2)$
	\item all these operators are self-adjoint in the sense of \eqref{eq_duality} \textit{only} for $D=2$ (in this case $f=1$).
\end{itemize}

\subsection{Different masses}
In this section we will present operators for banana diagrams in different mass case. As mentioned above, ansatz for $D=2$ differs from $D> 2$, so we will proceed with ansatz \eqref{g_i_init} with $a=0$. The computation complexity increases dramatically so we managed to compute the equation only for 2 loop case.

\subsubsection{$D=2$, 2 loops}
The Symanzik polynomials are
\begin{equation*}
	U=\alpha_1\alpha_2+\alpha_1\alpha_3+\alpha_2\alpha_3
\end{equation*}
\begin{equation}
	F=(\alpha_1\alpha_2+\alpha_1\alpha_3+\alpha_2\alpha_3)(m_1^2\alpha_1+m_2^2\alpha_2+m_3^2\alpha_3)-t\, \alpha_1\alpha_2\alpha_3
\end{equation}
Again, repeating \eqref{ansatz_cond_one_loop_d_2} we can derive the following operator:
\begin{equation}
	\begin{aligned}
		\Big\{&\Big(t((-m_1+m_2+m_3)^2-t)((m_1-m_2+m_3)^2-t)((m_1+m_2-m_3)^2-t)((m_1+m_2+m_3)^2-t)*\\
		&*((m_1^2-m_2^2)^2+(m_1^2-m_3^2)^2+(m_2^2-m_3^2)^2-(t-m_1^2)^2-(t-m_2^2)^2-(t-m_3^2)^2)\Big)\frac{d}{dt}^2+\\
		&+\Big(\left(m_1-m_2-m_3\right){}^3 \left(m_1+m_2-m_3\right){}^3 \left(m_1-m_2+m_3\right){}^3 \left(m_1+m_2+m_3\right){}^3-\\
		&-8 \left(m_1-m_2-m_3\right) \left(m_1+m_2-m_3\right) \left(m_1-m_2+m_3\right) \left(m_1+m_2+m_3\right)*\\
		&* \left(m_1^6-m_2^2 m_1^4-m_3^2 m_1^4-m_2^4
		m_1^2-m_3^4 m_1^2+10 m_2^2 m_3^2 m_1^2+m_2^6+m_3^6-m_2^2 m_3^4-m_2^4 m_3^2\right)t+\\
		&+(13 m_1^8-36 m_2^2 m_1^6-36 m_3^2 m_1^6+46 m_2^4 m_1^4+46 m_3^4 m_1^4-124 m_2^2 m_3^2 m_1^4-36 m_2^6 m_1^2-36 m_3^6 m_1^2-\\
		&-124 m_2^2 m_3^4 m_1^2-124
		m_2^4 m_3^2 m_1^2+13 m_2^8+13 m_3^8-36 m_2^2 m_3^6+46 m_2^4 m_3^4-36 m_2^6 m_3^2)t^2+\\
		&+8 \left(m_1^2+m_2^2+m_3^2\right) \left(m_1^4+6 m_2^2 m_1^2+6 m_3^2 m_1^2+m_2^4+m_3^4+6 m_2^2 m_3^2\right)t^3-\\
		&-37 (m_1^4-70 m_2^2 m_1^2-70 m_3^2 m_1^2-37 m_2^4-37 m_3^4-70 m_2^2 m_3^2)t^4+\\
		&+32 \left(m_1^2+m_2^2+m_3^2\right)t^5-9t^6\Big)\frac{d}{dt}-\\
		&-\left(m_1-m_2-m_3\right) \left(m_1+m_2-m_3\right) \left(m_1-m_2+m_3\right) \left(m_1+m_2+m_3\right)*\\
		&*\left(m_1^6-m_2^2 m_1^4-m_3^2
		m_1^4-m_2^4 m_1^2-m_3^4 m_1^2+6 m_2^2 m_3^2 m_1^2+m_2^6+m_3^6-m_2^2 m_3^4-m_2^4 m_3^2\right)+\\
		&+(5 m_1^8-8 m_2^2 m_1^6-8 m_3^2 m_1^6+6 m_2^4 m_1^4+6 m_3^4 m_1^4-8 m_2^2 m_3^2 m_1^4-8 m_2^6 m_1^2-8 m_3^6 m_1^2-8 m_2^2 m_3^4 m_1^2-\\
		&-8 m_2^4 m_3^2	m_1^2+5 m_2^8+5 m_3^8-8 m_2^2 m_3^6+6 m_2^4 m_3^4-8 m_2^6 m_3^2)t-\\
		&-2 \left(3 m_1^6-7 m_2^2 m_1^4-7 m_3^2 m_1^4-7 m_2^4 m_1^2-7 m_3^4 m_1^2+3 m_2^6+3 m_3^6-7 m_2^2 m_3^4-7 m_2^4 m_3^2\right)t^2-\\
		&-2 \left(m_1^4+8 m_2^2 m_1^2+8 m_3^2 m_1^2+m_2^4+m_3^4+8 m_2^2 m_3^2\right)t^3+\\
		&+7 \left(m_1^2+m_2^2+m_3^2\right)t^4-3t^5\Big\}B_3=0
		\end{aligned}
		\label{pf_2_loops_d_2_different}
\end{equation}
This operator is of the 2nd order as in the equal mass case which, in fact, will not be true for higher loop integrals. Moreover, it is \textit{self-adjoint} with
\begin{equation}
	f=\left((m_1^2-m_2^2)^2+(m_1^2-m_3^2)^2+(m_2^2-m_3^2)^2-(t-m_1^2)^2-(t-m_2^2)^2-(t-m_3^2)^2\right)^2
\end{equation}
Which equals non-singular multiplier of the leading coefficient squared.

\subsubsection{$D>2$, 2 loops}
Introducing the $6$-term symmetric combinations $M_{ijk} := m_1^{2i}m_2^{2j}m_3^{2k} + {\rm perms} $, at
generic $D$ the  equation is
\begin{equation}\label{N=3differentmasses3loops}
\begin{split}
    \left\{ c_4 \!\cdot\! \frac{\partial^4}{\partial t^4}
+c_3\frac{\partial^3}{\partial t^3}
+c_2\frac{\partial^2}{\partial t^2}+c_1\frac{\partial}{\partial t}+c_0\right\}
\frac{(\alpha_1\alpha_2+\alpha_2\alpha_3+\alpha_3\alpha_1)^{\frac{3(2-D)}{2}}}{\Omega_3^{3-D}}
=d \omega
\end{split}
\end{equation}
where:
\begin{equation}
\footnotesize
\begin{aligned}
    c_4 =& 8\left(t-(m_1+m_2+m_3)^2\right)\left(t-(m_1-m_2+m_3)^2\right)\left(t-(m_1+m_2-m_3)^2\right)\left(t-(m_1-m_2-m_3)^2\right) \times\\
    &\times\left((D-7)t^2+2(D-3)(m_1^2+m_2^2+m_3^2)t- m_1^4 + 2m_1^2m_2^2 - m_2^4 + 2m_1^2m_3^2 + 2m_2^2m_3^2 - m_3^4\right)
\end{aligned}
\end{equation}
and the rest of the coefficients (for a choice of $m_i$) are listed in Appendix \ref{AppendixA2}. Our results reproduce those of \cite{Adams:2015gva} and \cite{delaCruz:2024xit}. 

	\section{Fourier transform from coordinate to momentum space
	\label{sec:xtop}}

In this section we explain how to Fourier transform differential operators from position space to momentum space and back. This allows to convert the differential equations obtained in section \ref{sec:coord} into equation for actual loop integrals and compare them with the PF approach.

	\subsection{Momentum-space equations in $\Lambda$-formalism}
	Using the dilatation operator in $\Lambda$-formalism is very convenient for conversion from coordinate to momentum space. It consists of just three rules.
	Let $t:=p^2$.
	Then for the action on functions, which depend only on $x^2$ or $p^2=t$
	\be
	\hat\Box = \p_\mu\p_\mu = \p_\mu \frac{x_\mu}{|x|} \p= \frac{(D-1)}{|x|}\p + \p^2
	= \frac{1}{x^2} \left(\Lambda^2 + (D-2) \Lambda\right) \cong -t \nn \\
	\Lambda \cong -D-2t\frac{\p}{\p t} = -(D-2)-2\frac{\p}{\p t}t
	\nn \\
	x^2 \cong -\frac{\p}{\p p_\mu}\frac{\p}{\p p_\mu} = -\frac{\p}{\p p_\mu}2p_\mu\frac{\p}{\p t}
	= -2D\frac{\p}{\p t} - 4t\frac{\p^2}{\p t^2} = 2\Lambda \frac{\p}{\p t}
	\label{momrep}
	\ee
	We remind that $\p:=\frac{\p}{\p x}$ and $\cong$ denotes Fourier transform of an operator.
	To understand these equations one can keep in mind that
	\be
	\frac{\p}{\p p_\mu} = 2p_\mu \frac{\p}{\p t}
	\ \ \ \ \Longrightarrow \ \ \ \
	2t\frac{\p}{\p t} = p_\mu\frac{\p}{\p p_\mu} \cong -\frac{\p}{\p x_\mu} x_\mu = -(D+\Lambda)
	\ee
	\\\\
	With the help of  these transition formulas (\ref{momrep})  one can transform equation in position space to equations in momentum space. Let us explain in detail how that works, while also illustrating the derivation of \eqref{momrep}. Consider then Fourier transform of a Lorenz invariant function:
	\begin{equation}
		f(x) = \int d^D p \, e^{i p_\mu x^\mu}\tilde{f}(t)
	\end{equation}
	Now suppose that one has a simple equation:
	\begin{equation}
		\left(\Lambda +  \nu \right) f(x)= 0 \ \   \Rightarrow \ \ f(x) = \dfrac{c_1}{x^{\nu}}
	\end{equation}
	Then we write:
	\begin{equation}
		\begin{split}
			&\left(\Lambda +  \nu \right) \int d^D p \, e^{i p_\mu x^\mu}\tilde{f}(t)  = \int d^D p  \left(i x^\mu p_\mu  +\nu \right) e^{i  p_\mu x^\mu } \tilde{f}(t) =
			\\
			&= \int d^Dp \, e^{i  p_\mu x^\mu } \left( - p_\mu \dfrac{\partial}{\partial p_\mu} - D + \nu \right)\tilde{f}(t) = \int d^Dp \, e^{i  p_\mu x^\mu } \left( -2t \dfrac{\partial}{\partial t} - D + \nu \right)\tilde{f}(t)
		\end{split}
	\end{equation}
	Therefore for the Fourier transform we obtain an equation:
	\begin{equation}
		\left( -2t \dfrac{\partial}{\partial t} - D + \nu \right)\tilde{f}(t) = 0
	\end{equation}
	which is solve by $\tilde{f}(t) = \dfrac{c_2}{t^{\frac{D-\nu}{2}}}$ which is indeed the Fourier transform of the initial function.
	\\\\
	One should be cautious, however, when comparing solutions of the position and momentum space equation as Fourier transforms of smooth functions may lead to distribution. We will not discuss this in detail in the examples of real equations for diagrams, however let us provide a simple illustration at $D=1$.
	Consider:
	\begin{equation}
		x^2 \dfrac{\partial}{\partial x } f(x)=0
	\end{equation}
	The F.T. of this equation would lead to:
	\begin{equation}
		\dfrac{\partial^2}{\partial p^2}\left(p  \tilde{f}(p) \right) =0
	\end{equation}
	The apparent mismatch of the number of solutions is due to the fact that we would only count regular ones. However by formally allowing $x \delta(x)=0$ we can match solutions to both equations in the following way:
	\begin{center}
		\begin{tabular}{c|c}
			$x^2 \dfrac{\partial}{\partial x } f(x)$ &
			$\dfrac{\partial^2}{\partial p^2}\left(p  \tilde{f}(p) \right)$
			\\
			\hline
			$\delta(x)$ & $1$
			\\
			$\theta(x)$ & $\frac{1}{p}$
			\\
			$ 1$ & $\delta(p)$
		\end{tabular}
	\end{center}
	We will later see, that the order of differential operators for banana diagrams in momentum and position space is different, furthermore, they don't match the ones obtained through the Picard-Fuchs approach, which we review in the next section. The careful treatment of this mismatch is an important question for further research.
	\\\\
	Going back to equations for banana graphs, let us provide a prime example of using the Fourier transform to obtain the differential equation for the momentum space banana integral. The unequal mass generic dimension equation for the one loop graph
	\begin{equation}
		\begin{split}
			\left\{x^2 \Big(\Box+(m_1+m_2)^2\Big) \right.  &\Big(\Box + (m_1-m_2)^2\Big)
			+ \\
			&\left. +
			2(D-1)\Lambda(\Box+m_1^2+m_2^2)
			+ 2(D-1)(D-2)(\Box+m_1^2+m_2^2)\right\}f(x) = 0
			\label{1loopcoordnate}
		\end{split}
	\end{equation}
	
	turns into the equation
	\begin{equation}
	    \begin{split}
	        -2\frac{\p}{\p t} \Big\{ 2\frac{\p}{\p t} t(t-(m_1-m_2)^2)&(t-(m_1+m_2)^2)
	- 
 \\
 &-(D+2)t^2+6(m_1^2+m_2^2)t +(D-4)(m_1^2-m_2^2)^2
	\Big\} \tilde f(t) = 0
	    \end{split}
	\end{equation}
	\label{n=2trough_t}

	The second factor is of the first order in $\frac{\p}{\p t}$ --
	this matches the expectation about Picard-Fuchs equation
	for the corresponding period.
	Note that at $D=1$ the three last terms combine into
	the same structure $-3(t-(m_1-m_2)^2)(t-(m_1+m_2)^2)$
	which appears in the first term.

	\paragraph{Additional degree introduces by $\Lambda$}
	and when the same transformation is applied do the whole equation \eqref{1loopcoord}, we are left with the Fourier integral of the expression
	\[\begin{split}  & 4 (m_1^2 - m_2^2) \Big( - 2 D  \frac{\partial}{\partial t} - 4 t \frac{\partial^2}{\partial t^2}\Big) \Big( 16 t^2 \big(t^2 - 2 t (m_1^2+m_2^2) + (m_1^2-m_2^2)^2\big) \frac{\partial^4 \tilde{f}}{\partial t^4} +\\
		& \qquad 16 t \big(11 t^2 - (D+13) (m_1^2+m_2^2) t + (D+2) (m_1^2+m_2^2)\big) \frac{\partial^3 \tilde{f}}{\partial t^3} -\\
		&\qquad - 4 \big((D^2 -4 D -120) t^2 + 6 (3D+10) (m_1^2+m_2^2) t - D (D+2) (m_1^2-m_2^2)^2\big) \frac{\partial^2 \tilde{f}}{\partial t^2} - \\
		& \qquad - 16 \big((D^2 -4D -18) t +3 D(m_1^2 +m_2^2)\big) \frac{\partial \tilde{f}}{\partial t} - 8 (D-4) t \tilde{f}(t) \Big)
	\end{split}\]
	which therefore has to vanish and constitutes the momentum--space equation on one--loop banana.

	\paragraph{Determinant in momentum space}
	One can make the Fourier transform directly in the determinant \eqref{eq:DeterminantEq} which enables us to write a general form of the equation in momentum space. Note that doing so would not introduce any issues of taking determinant of non-commuting operators. This is due to the fact that all the coefficients $C_{n,\vec{\epsilon}}$ in \eqref{eq:DeterminantEq} are made out of the multiplication by $x^2$ operators, which may look more complicated in momentum space, but are still commuting.
	To get an explicit expression for the coefficients of this operator we may use the simple transition formula:
	\begin{equation}
		x^{2i}= (-2)^i \sum_{k=0}^i 2^k \binom{i}{k} \left( \prod_{l=1}^{i-k}(D+2i-2l) \right) t^k\dfrac{d^{i+k}}{dt^{i+k}}
	\end{equation}
	\subsection{Inverse transformation -- from momenta to coordinates}
	As we have seen, the Picard-Fuchs equations are obtained independently of the position space ones. Thus, it is also interesting to see, what appears in coordinate space if we Fourier transform the PF operator.	Since $d/dt$ does not allow immediate transform into the $x$-space, one has to first rewrite the differential equation using the momentum space Euler/dilation operator:
	\begin{equation}
		\Theta=t\frac{\partial}{\partial t}
	\end{equation}
	In general having some operator in the form
	\begin{equation}
		P=\sum_i a_i(t)\frac{\partial^i}{\partial t^i}
	\end{equation}
	We transform it to Euler form by multiplying by a certain degree in $t$ from the left
	\begin{equation}
		P=\sum_i b_i(t) \Theta^i
	\end{equation}
	Then the inverse Fourier transform is straightforward and is determined by the following rules
	\begin{equation}
	    \begin{split}
     t \cong \p^2-\frac{(D-1)}{|x|}\p  =&  \frac{1}{x^2}\left(\Lambda^2 + (D-2)\Lambda \right)  \\
	\Theta=-t\frac{\p}{\p t} \cong&\frac{1}{2}(D+\Lambda)
	\label{momentum_to_coord_var_change}
	    \end{split}
	\end{equation}

\section{Comparing position space to Picard-Fuchs and factorization
\label{sec:factorization}}
So far we have presented 2 approaches of computing nullifying operators for banana Feynman integrals. The coordinate approach seems to have two advantages: it only requires very basic ''first principle'' considerations to construct and is also quite fast to produce equation.
On the other hand, PF operators uncover the underlying geometry and are minimal already in momentum space. On the other hand the position space approach seems to produce much more complicated operators when different masses are introduced.
\\

The approaches can be compared by Fourier transforming the coordinate space equations. As motivated in the introduction, one would expect the kernel of the operator $\hat{E}_{t}^{n}$ to at least intersect with that of $\widehat{PF}$. A more optimistic expectation would be for one kernel to lie inside the other. There is no \emph{a priori} reason, for the operators to coincide exactly. The results of \cite{Mishnyakov:2023wpd} of indicate that for equal masses, however, this is exactly the case. On the other hand as already outlined in \cite{Mishnyakov:2023sly} this is not the case when masses become different. We explore this relation deeper in this section. Mainly we demonstrate that in the different mass case, the order of $\hat{E}_{t}^{(n)}$ is much larger than that of the PF operator, i.e., it is not minimal. However, we observe that they are related by factorization:
\begin{equation}
	 \hat{E}^{(n)}_t  =  \hat{\bf{B}} \cdot \widehat{PF}
\end{equation}
hence the kernel of the Picard-Fuchs operator is contained within the Fourier transform of the position space operator. Furthermore, one can also do the inverse, namely having a Picard-Fuchs equation translate it to position space, to obtain an operator $\widehat{PF}_x$. What we observe is that it now becomes non-minimal in the $\partial_x$ derivative and its factorization is the $\mathcal{E}_x$ operators.
\\\\
The relation can be summarized in the following table:

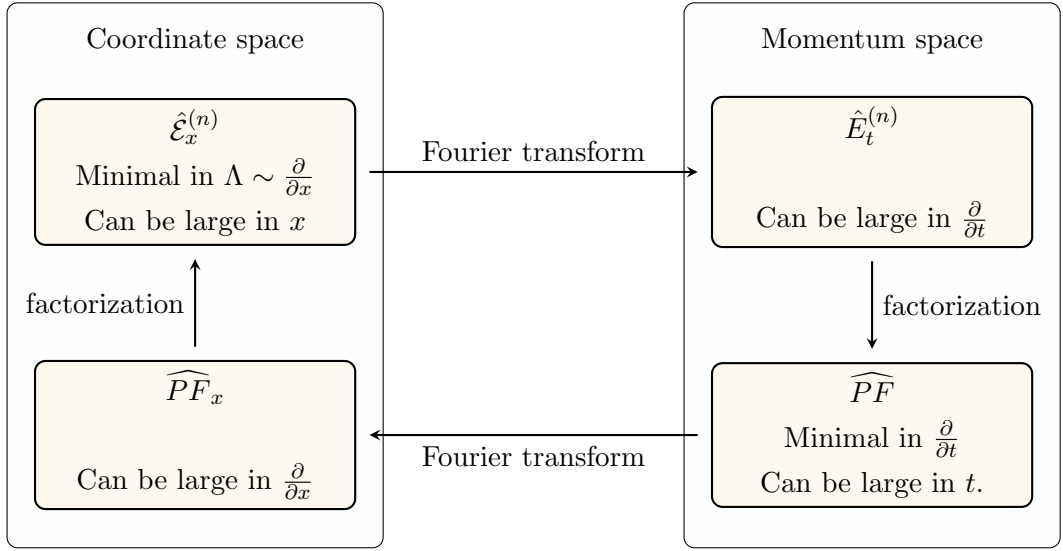
\begin{center}
	\begin{tikzpicture}[shorten <=5pt,shorten >=5pt,>=stealth,style=thick]
		\node[fill=champagne!30,draw,rounded corners] (A) at  (-5,2) {\begin{minipage}{4cm}
				\centering
				$\hat{\mathcal{E}}_x^{(n)}$
				\\
				Minimal in $\Lambda \sim \frac{\partial}{\partial x}$
				\\
				Can be large in $x$
		\end{minipage}};
		\node[fill=champagne!30,draw,rounded corners] (B) at  (4,2) {\begin{minipage}{4cm}
				\centering
				$\hat{E}_t^{(n)}$				\\
				\hfill
				\\Can be large in $\frac{\partial}{\partial t}$
		\end{minipage}};
		\node[fill=champagne!30,draw,rounded corners] (C) at  (4,-1.5) {\begin{minipage}{4cm}
				\centering
				$\widehat{PF}$				\\
				Minimal in $\frac{\partial}{\partial t}$\\
				Can be large in $t$.
		\end{minipage}};
		\node[fill=champagne!30,draw,rounded corners] (D) at  (-5,-1.5) {\begin{minipage}{4cm}
			\centering
			$\widehat{PF}_x$				\\
			\hfill
			\\
			Can be large in $\frac{\partial}{\partial x}$
	\end{minipage}};
	\node (AA) at (-5,3.75) {Coordinate space}; 
	\begin{scope}[on background layer]
	\draw [fill=mintcream!30,rounded corners] (-7.5,4.25) rectangle (-2.5,-3);
	\end{scope}
\node (AA) at (4,3.75) {Momentum space}; 
\begin{scope}[on background layer]
	\draw [fill=mintcream!30,rounded corners] (6.5,4.25) rectangle (1.5,-3);
\end{scope}
	\draw [->] (A)-- node[midway, above] {Fourier transform}  (B);
	\draw [->] (C) -- node[midway, below] {Fourier transform}  (D);
	\draw [->] (B) -- node[midway, right] {factorization}  (C);
	\draw [->] (D) -- node[midway, left] {factorization}  (A);
	\end{tikzpicture}
\end{center}

As seen from the Fourier transform, the order of the differential operator in momentum space is governed by the degree in $x$ in position space and vice versa. Therefore, it is convenient to think of them in terms of a bi-grading: in positions space in powers of $x$ and $\partial_x$ and in momentum space respectively $t$ and $\partial_t$. Therefore, it is useful to introduce a bi-grading both in position and momentum space:
\begin{equation}
(\text{degree in } x  \  (\text{ resp. } t) \, , ,\ \text{ degree in } \partial_x \   (\text{resp. } \partial_t ) )
\end{equation}
\subsection{Equal mass}
\label{eq_mass_fact}

As was observed in \cite{Mishnyakov:2023wpd} at equal mass the operators $E_t$ and $\widehat{PF}_t$ are essentially the same. A slight modification is due in the present paper because of the use of the $\Lambda$ operator. It introduces additional powers of $x$ from the left in the operator, which translates to more derivatives in momentum space. However, it always appears in the form of an additional $x^2$ factor, hence can be simply accounted for. Therefore, the diagram is the following:
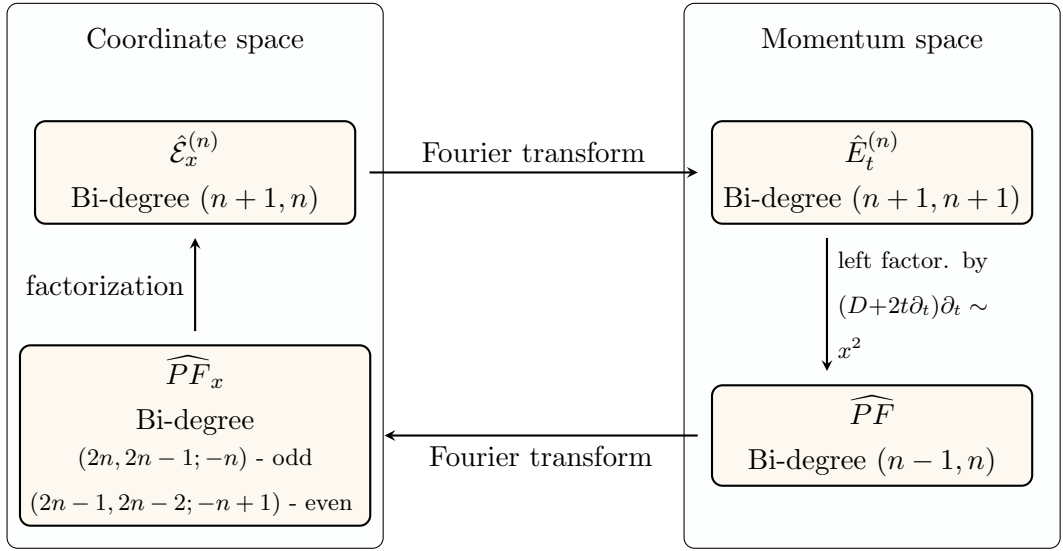
\begin{center}
	\begin{tikzpicture}[shorten <=5pt,shorten >=5pt,>=stealth,style=thick]
		\node[fill=champagne!30,draw,rounded corners] (A) at  (-5,2) {\begin{minipage}{4cm}
				\centering
				$\hat{\mathcal{E}}_x^{(n)}$
				\\
				Bi-degree $(n+1,n)$
		\end{minipage}};
		\node[fill=champagne!30,draw,rounded corners] (B) at  (4,2) {\begin{minipage}{4cm}
				\centering
				$\hat{E}_t^{(n)}$				\\
				Bi-degree $(n+1,n+1)$
		\end{minipage}};
		\node[fill=champagne!30,draw,rounded corners] (C) at  (4,-1.5) {\begin{minipage}{4cm}
				\centering
				$\widehat{PF}$				\\
				Bi-degree $(n-1,n)$
		\end{minipage}};
		\node[fill=champagne!30,draw,rounded corners] (D) at  (-5,-1.5) {\begin{minipage}{4.4cm}
				\centering
				$\widehat{PF}_x$				\\
				Bi-degree \\{\footnotesize 
                $(2n, 2n-1;-n)$ - odd
                \\$(2n-1,2n-2;-n+1)$ - even
                }
                
		\end{minipage}};
		\node (AA) at (-5,3.75) {Coordinate space};
		\begin{scope}[on background layer]
			\draw [fill=mintcream!30,rounded corners] (-7.5,4.25) rectangle (-2.5,-3);
		\end{scope}
		\node (AA) at (4,3.75) {Momentum space}; 
		\begin{scope}[on background layer]
			\draw [fill=mintcream!30,rounded corners] (6.5,4.25) rectangle (1.5,-3);
		\end{scope}
		\draw [->] (A)-- node[midway, above] {Fourier transform}  (B);
		\draw [->] (C) -- node[midway, below] {Fourier transform}  (D);
		\draw [->] ([xshift=-0.6cm]B.south) -- node[midway, right] {\begin{minipage}{2cm}
				{\footnotesize left factor. by $(D+2t{\partial_t})\partial_t 
\sim x^2$}
			\end{minipage}
		}  ([xshift=-0.6cm]C.north);
		\draw [->] (D) -- node[midway, left] {factorization}  (A);
	\end{tikzpicture}
\end{center}

By the notation $(m,l;-k)$ or, below, $(m,l(-k))$ we mean that the operator has order $m$ and the leading coefficient is rational function with numerator of degree $l$ and denominator of degree $k$.

\paragraph{Computation complexity remark}
As one can see from the diagram above there are 2 ways to get $\widehat{PF}$ equation:
\begin{enumerate}
    \item straightforward computation of $\widehat{PF}$
    \item compute $\hat{\mathcal{E}}_x^{(n)} \xrightarrow{Fourier} \hat{E}_t^{(n)} \xrightarrow{\text{factorization}} \widehat{PF}$
\end{enumerate}
Both of the ways lead to the same answer, however each of them have their own drawbacks and advantages. \\
The \textbf{first} way allows computing equal mass operator for any diagram but its computational complexity grows exponentially with the number of loops (edges).\\
The \textbf{second} way starts with the much easier computable $\hat{\mathcal{E}}_x^{(n)}$ but finishes difficult factorization process, as the operator $\hat{E}_t^{(n)}$ is not Fuchsian and the factorization is not unique. However, in the case of equal masses the left factor does not change depending on dimension or number of loops, hence making this way much easier.\\
To compare the complexity difference, we present two tables with computation time for different loops
\begin{table}[h]
    \begin{minipage}{.5\linewidth}
      \centering
        \begin{tabular}{c|c}
            loops & time  \\
            \hline
            1 & $<1$ sec.\\
            \dots & \dots \\
            7 & several hours \\
            \dots & \dots \\
            30 & many hours
        \end{tabular}
        \caption{The first way}
    \end{minipage}%
    \begin{minipage}{.5\linewidth}
      \centering
        \begin{tabular}{c|c}
            loops & time  \\
            \hline
            1 & $<1$ sec.\\
            \dots & \dots \\
            7 & several seconds\\
            \dots & \dots \\
            30 & 2-3 hours
        \end{tabular}
        \caption{The second way}
    \end{minipage} 
\end{table}
We do not present the result for $30$ loops as it becomes enormously large and there is still no compact language to express it.
\paragraph{Examples}
Moving further, we spell out a few examples explicitly.\\
At 1-loop we have \eqref{eq_mass_coord_1_loop} :
\begin{equation}
		\mathcal{E}^{(2)}_x  = 	\left(\Lambda ^3+3 (D-2) \Lambda ^2+  \left(2 (D-2)^2+4 m^2 x^2\right) \Lambda +4 (D-1) m^2 x^2 \right)
\end{equation}
which after Fourier transform is in momentum space
\begin{equation}
	\hat{E}^{(2)}_t = \left(D+2t\frac{\p}{\p t}\right)\frac{\p}{\p t} \cdot \widehat{PF}
\end{equation}
and the Picard-Fuchs operator is given by \eqref{PF1loopequalmasses}:
\begin{equation}
	\widehat{PF} = t(t-4m^2)\frac{\p}{\p t} -(D-4)t -4m^2 
\end{equation}
To make it suitable for Fourier transform we rewrite it as:
\begin{equation}
	\widehat{PF} = (t-4m^2)\Theta -(D-4)t -4m^2 
\end{equation}

which then produces:
\begin{equation}
    \widehat{PF}_x=\frac{1}{x^2}\left(\Lambda ^3+3 (D-2) \Lambda ^2+  \left(2 (D-2)^2+4 m^2 x^2\right) \Lambda +4 (D-1) m^2 x^2 \right)=\frac{1}{x^2}\mathcal{E}^{(2)}_x
\end{equation}

Hence, the diagram in this case looks like:
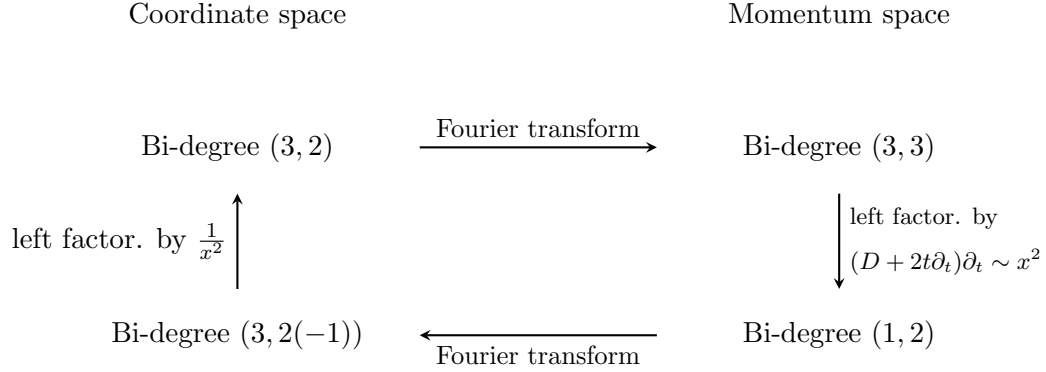
\begin{center}
	\begin{tikzpicture}[shorten <=8pt,shorten >=8pt,>=stealth,style=thick]
		\node (A) at  (-4,2) {\begin{minipage}{4cm}
				\centering
				 Bi-degree $(3,2)$
		\end{minipage}};
		\node (B) at  (4,2) {\begin{minipage}{4cm}
				\centering
			 Bi-degree $(3,3)$
		\end{minipage}};
		\node (C) at  (4,-0.5) {\begin{minipage}{4cm}
				\centering
				Bi-degree $(1,2)$
		\end{minipage}};
		\node (D) at  (-4,-0.5) {\begin{minipage}{4cm}
				\centering
				Bi-degree $(3,2(-1))$
		\end{minipage}};
		\node (AA) at (-4,3.75) {Coordinate space}; 
		\node (AA) at (4,3.75) {Momentum space}; 
		\draw [->] (A)-- node[midway, above] {{\small Fourier transform}}  (B);
		\draw [->] (C) -- node[midway, below] {{\small Fourier transform}}  (D);
		\draw [->] (B) -- node[midway, right] {\begin{minipage}{3cm}
				{\footnotesize left factor. by\\ $(D+2t{\partial_t})\partial_t \sim x^2$}
			\end{minipage}
		}  (C);
		\draw [->] (D) -- node[midway, left] {left factor. by $\frac{1}{x^2}$ }  (A);
	\end{tikzpicture}
\end{center}
At two loops:
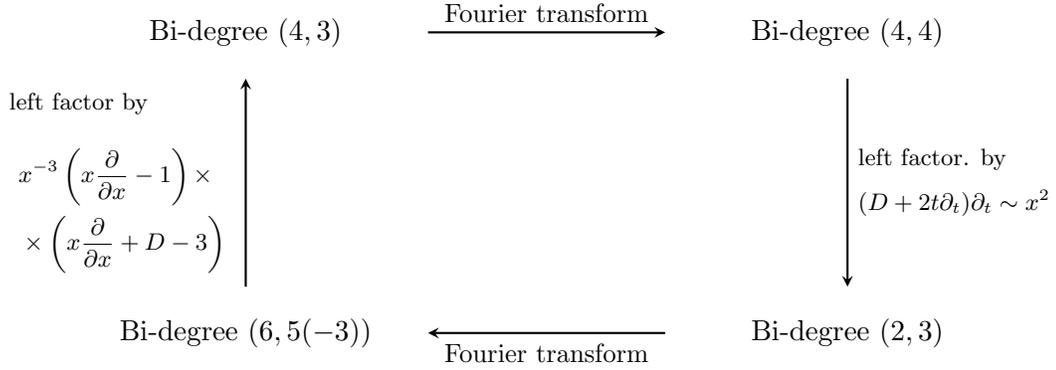
\begin{center}
	\begin{tikzpicture}[shorten <=8pt,shorten >=8pt,>=stealth,style=thick]
		\node (A) at  (-4,2) {\begin{minipage}{4cm}
				\centering
				 Bi-degree $(4,3)$
		\end{minipage}};
		\node (B) at  (4,2) {\begin{minipage}{4cm}
				\centering
			 Bi-degree $(4,4)$
		\end{minipage}};
		\node (C) at  (4,-1.5-0.5) {\begin{minipage}{4cm}
				\centering
				Bi-degree $(2,3)$
		\end{minipage}};
		\node (D) at  (-4,-1.5-0.5) {\begin{minipage}{4cm}
				\centering
				Bi-degree $(6,5(-3))$
		\end{minipage}};
		\draw [->] (A)-- node[midway, above] {{\small Fourier transform}}  (B);
		\draw [->] (C) -- node[midway, below] {{\small Fourier transform}}  (D);
		\draw [->] (B) -- node[midway, right] {\begin{minipage}{3cm}
				{\footnotesize left factor. by\\ $(D+2t{\partial_t})\partial_t \sim x^2$}
			\end{minipage}
		}  (C);
		\draw [->] (D) -- node[midway, left] {\begin{minipage}{3cm}
		   {\footnotesize  left factor by 
     \\
     \begin{equation*}
         \begin{split}
              &x^{-3}\left(x\frac{\p}{\p x}-1\right)\times
              \\
              &\times\left(x\frac{\p}{\p x}+D-3\right)
         \end{split}
     \end{equation*}
    } 
		\end{minipage}
  }  (A);
	\end{tikzpicture}
\end{center}

\subsection{Different masses}
The different masses case is more complicated.  There is also a statement of \cite{delaCruz:2024xit} about the order of the Picard-Fuchs operator.

\paragraph{Examples.}
At one loop we have \eqref{1loopcoord}:
\begin{equation}
\begin{aligned}
    \mathcal{E}^{(2)}_x =& \Lambda^4+2(D-5)\Lambda^3+(2 (m_1^2 + m_2^2) x^2 + D (5 D-26) + 32)\Lambda^2 + \\
    &+ (2 (2D-3) (m_1^2 + m_2^2) x^2 + 2 (D-4) (D-2)^2)\Lambda + \\
    &+ (m_1^2 - m_2^2)^2 x^4 +  2 (D-2) (D-1) (m_1^2 + m_2^2) x^2
\end{aligned}
\end{equation}
which after Fourier transform is in momentum space
\begin{equation}
	\hat{E}^{(2)}_t = \left(D+2t\frac{d}{dt}\right)\frac{d^2}{dt^2}\widehat{PF}
\end{equation}
and the Picard-Fuchs operator is given by \eqref{PF1loopdiffmasses}:
\begin{equation}
	\widehat{PF} =2t\big(t-(m_1-m_2)^2\big)\big(t-(m_1+m_2)^2\big)\frac{\p}{\p t} -
(D-4)t^2 -2(m_1^2+m_2^2)t +(D-2)(m_1^2-m_2^2)^2
\end{equation}
To make it suitable for Fourier transform we rewrite it as:
\begin{equation}
	\widehat{PF} = 2\big(t-(m_1-m_2)^2\big)\big(t-(m_1+m_2)^2\big)\Theta -
(D-4)t^2 -2(m_1^2+m_2^2)t +(D-2)(m_1^2-m_2^2)^2
\end{equation}
which then produces:
\begin{equation}
    \widehat{PF}_x=\frac{1}{x^3}\left(\Lambda-1\right)\mathcal{E}^{(2)}_x
\end{equation}
Hence, the diagram in this case looks like:
\begin{center}
	\begin{tikzpicture}[shorten <=8pt,shorten >=8pt,>=stealth,style=thick]
		\node (A) at  (-4,2) {\begin{minipage}{4cm}
				\centering
				 Bi-degree $(4,3)$
		\end{minipage}};
		\node (B) at  (4,2) {\begin{minipage}{4cm}
				\centering
			 Bi-degree $(4,4)$
		\end{minipage}};
		\node (C) at  (4,-1) {\begin{minipage}{4cm}
				\centering
				Bi-degree $(1,3)$
		\end{minipage}};
		\node (D) at  (-4,-1) {\begin{minipage}{4cm}
				\centering
				Bi-degree $(5,4(-3))$
		\end{minipage}};
		\draw [->] (A)-- node[midway, above] {{\small Fourier transform}}  (B);
		\draw [->] (C) -- node[midway, below] {{\small Fourier transform}}  (D);
		\draw [->] (B) -- node[midway, right] {\begin{minipage}{3cm}
				{\footnotesize left factor. by\\ $(D+2t{\partial_t})\partial_t^2$}
			\end{minipage}
		}  (C);
		\draw [->] (D) -- node[midway, left] {\begin{minipage}{3cm}
		   {\footnotesize  left factor by 
     \\
     \begin{equation*}
         \begin{split}
              &x^{-3}\left(x\frac{\p}{\p x}-1\right)
         \end{split}
     \end{equation*}
    } 
		\end{minipage}
  }  (A);
	\end{tikzpicture}
\end{center}
and at two loops
\begin{center}
	\begin{tikzpicture}[shorten <=8pt,shorten >=8pt,>=stealth,style=thick]
		\node (A) at  (-4,2) {\begin{minipage}{4cm}
				\centering
				 Bi-degree $(8,11)$
		\end{minipage}};
		\node (B) at  (4,2) {\begin{minipage}{4cm}
				\centering
			 Bi-degree $(12,10)$
		\end{minipage}};
		\node (C) at  (4,-0.5) {\begin{minipage}{4cm}
				\centering
				Bi-degree $(4,9)$
		\end{minipage}};
		\node (D) at  (-4,-0.5) {\begin{minipage}{4cm}
				\centering
				Bi-degree $(16,15(-11))$
		\end{minipage}};
		\node (AA) at (-4,3.75) {Coordinate space}; 
		\node (AA) at (4,3.75) {Momentum space}; 
		\draw [->] (A)-- node[midway, above] {{\small Fourier transform}}  (B);
		\draw [->] (C) -- node[midway, below] {{\small Fourier transform}}  (D);
		\draw [->] (B) -- node[midway, right] {factorization		}  (C);
		\draw [->] (D) -- node[midway, left] {factorization    }  (A);
	\end{tikzpicture}
\end{center}
To illustrate the special effect of $D=2$ in momentum space consider also the two loop example with $D=2$:
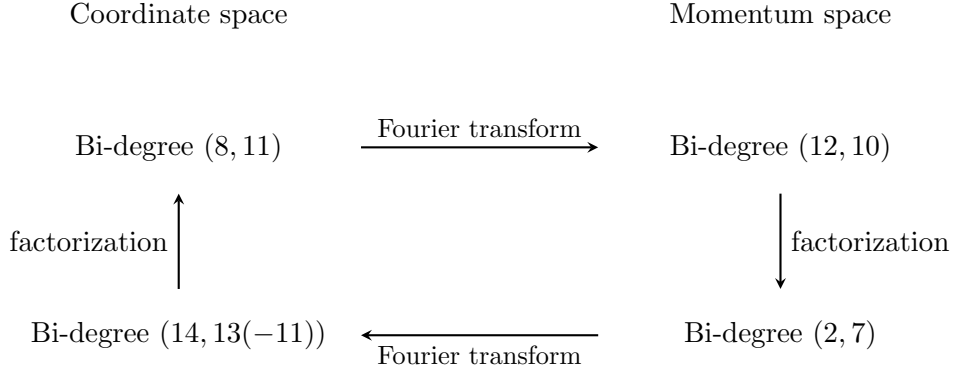
\begin{center}
	\begin{tikzpicture}[shorten <=8pt,shorten >=8pt,>=stealth,style=thick]
		\node (A) at  (-4,2) {\begin{minipage}{4cm}
				\centering
				 Bi-degree $(8,11)$
		\end{minipage}};
		\node (B) at  (4,2) {\begin{minipage}{4cm}
				\centering
			 Bi-degree $(12,10)$
		\end{minipage}};
		\node (C) at  (4,-0.5) {\begin{minipage}{4cm}
				\centering
				Bi-degree $(2,7)$
		\end{minipage}};
		\node (D) at  (-4,-0.5) {\begin{minipage}{4cm}
				\centering
				Bi-degree $(14,13(-11))$
		\end{minipage}};
		\node (AA) at (-4,3.75) {Coordinate space}; 
		\node (AA) at (4,3.75) {Momentum space}; 
		\draw [->] (A)-- node[midway, above] {{\small Fourier transform}}  (B);
		\draw [->] (C) -- node[midway, below] {{\small Fourier transform}}  (D);
		\draw [->] (B) -- node[midway, right] {factorization		}  (C);
		\draw [->] (D) -- node[midway, left] {factorization    }  (A);
	\end{tikzpicture}
\end{center}

\subsection{Comparing (the number of) solutions}\label{sec:numberofsolutions}
\label{number_of_solutions}
In this section we expand a little on the issue of comparing the number of solutions that was raised in sec. \ref{sec:xtop}. It is clear that this is necessary to properly understand the relation between position and momentum space operators. However, we do not intend to give an extensive answer.
It is clear that the order of differential equation in position and momentum spaces will generically be different, as the degree in $x$ and $\partial_x$ are not related. This means that counting has to be done with care.
\\

A textbook example of this phenomenon is provided by the Airy equation:
\begin{equation}
	\left(\frac{\p^2}{\p x^2}-x\right)g(x)=0
\end{equation}
It has two linear independent solutions, however taking Fourier leads to first order linear differential equation with only one solution:
\begin{equation}
	\left(i\frac{\p}{\p p} + p^2\right)g(p)=0\ \rightarrow\ g(p)=e^{i p^3/3 }
\end{equation}
where:
\begin{equation}
    g(x) = \int g(p) e^{ip x } = \int e^{ip^3/3}e^{ip x }
\end{equation}
The two initial solutions emerge since we are forced to choose an integration contour in the inverse transform, such that the integral converge. There are exactly 3 (homology classes of) such contours $\Gamma_1, \Gamma_2, \Gamma_3$ which satisfy this condition. However, only 2 of them are independent. Integrating over each of the two contours, we get the desired coordinate space solutions.
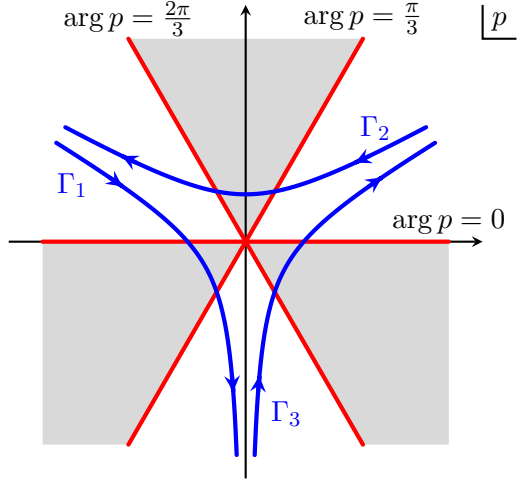
\begin{figure}[H]
	\centering
	\begin{tikzpicture}[scale=0.9]
        \begin{scope}[rotate around z=0]
		\fill[opacity=0.3, gray] (0, 0) -- ({ 3/tan(180/3) }, 3) -- ({ -3/tan(180/3) }, 3)  -- (0,0);
		\fill[opacity=0.3, gray] (0, 0) -- ({ -3/tan(180/3) }, -3) -- (-3, -3) -- (-3,0) -- (0,0);
		\fill[opacity=0.3, gray] (0, 0) -- (3, 0) -- (3,-3) -- ({ 3/tan(180/3) }, -3) -- (0,0);
        \end{scope}
		\begin{scope}[thick]
			\draw[-stealth] (0,-3.5) -- (0,3.5);
			\draw[-stealth] (-3.5,0) -- (3.5,0);
			\draw (3.5, 3.5) -- (3.5, 3) -- (4, 3);
			\draw (3.75, 3.25) node {$p$};
		\end{scope}
		\begin{scope}[ultra thick, cap=round]
    		\draw[red]  ({ -3/tan(180/3) }, 3) -- ({ 3/tan(180/3) }, -3) ;
                \draw ({ -3/tan(180/3) },3.3) node {$ \arg p = \frac{2\pi}{3}$};
                
			\draw[red] ({ 3/tan(180/3) }, 3) -- ({ -3/tan(180/3) }, -3);
                \draw ({ 3/tan(180/3) }, 3.3) node {$ \arg p = \frac{\pi}{3}$};
			
			\draw[red] (3, 0) -- (3/2, 0) -- (-3, 0);
                \draw (3,0.3) node {$ \arg p = 0$}; 
		\end{scope}
		\begin{scope}[ultra thick, cap=round,rotate around z=120+30]
			\draw[domain=-1.53:1.53, smooth, variable=\t, blue] plot ({-0.7*cosh(\t)}, {-1.73*0.7*sinh(\t)});
			\draw[domain=-1.2:-1.1, smooth, variable=\t, blue, -stealth] plot ({-0.7*cosh(\t)}, {-1.73*0.7*sinh(\t)});
			\draw[domain=1.1:1.2, smooth, variable=\t, blue, -stealth] plot ({-0.7*cosh(\t)}, {-1.73*0.7*sinh(\t)});
		\end{scope}
		\begin{scope}[ultra thick, cap=round,rotate around z=30]
			\draw[domain=-1.53:1.53, smooth, variable=\t, blue] plot ({-0.7*cosh(\t)}, {-1.73*0.7*sinh(\t)});
			\draw[domain=-1.2:-1.1, smooth, variable=\t, blue, -stealth] plot ({-0.7*cosh(\t)}, {-1.73*0.7*sinh(\t)});
			\draw[domain=1.1:1.2, smooth, variable=\t, blue, -stealth] plot ({-0.7*cosh(\t)}, {-1.73*0.7*sinh(\t)});
		\end{scope}
		
		\begin{scope}[ultra thick, cap=round,rotate around z=240+30]
			\draw[domain=-1.53:1.53, smooth, variable=\t, blue] plot ({-0.7*cosh(\t)}, {-1.73*0.7*sinh(\t)});
			\draw[domain=-1.2:-1.1, smooth, variable=\t, blue, -stealth] plot ({-0.7*cosh(\t)}, {-1.73*0.7*sinh(\t)});
			\draw[domain=1.1:1.2, smooth, variable=\t, blue, -stealth] plot ({-0.7*cosh(\t)}, {-1.73*0.7*sinh(\t)});
		\end{scope}
		
		\begin{scope}[ultra thick, cap=round,rotate around z=30]
			\draw (-1.8,2) node {\textcolor{blue}{$\Gamma_1$}};
			\draw (2.5,0.5) node {\textcolor{blue}{$\Gamma_2$}};
			\draw (-0.75,-2.5) node {\textcolor{blue}{$\Gamma_3$}};
		\end{scope}
	\end{tikzpicture}
\end{figure}

\paragraph{Toy cases for Feynman integrals.} We believe that similar phenomena happen for the Picard-Fuchs and position space equations.
When moving on to Feynman integrals, we encounter additional complications compared to the Airy example. These are mostly due to the necessity to include distributional solutions.

Consider the toy case $D=1$. Then the momentum space equation of motion are given by:
\begin{equation}
    (p^2-m^2)G(p)=0
\end{equation}
which has a two-dimensional solution space
\begin{equation}
	G(p) \sim \mathbb{C}\delta(p+m) + \mathbb{C}\delta(p-m)
\end{equation}
The maximal cut propagator is a special combination of these two:
\begin{equation}
	 \delta(p^2-m^2)=\frac{\delta(p+m)+\delta(p-m)}{2m}
\end{equation}
If we suppose that $m>0$ two possible inverse transforms are written as:
\begin{equation}
    e^{i m x} = \int_{-\infty}^{\infty} \delta(p-m) e^{i p x} \, \quad e^{-i m x} = \int_{-\infty}^{\infty} \delta(p+m) e^{i p x}
\end{equation}
The one loop functions are then integrals of the sort:
\begin{equation}
    B_2(p) \sim \int_\Gamma \delta(p-q \pm m)\delta(q \pm m) 
\end{equation}
Depending on the choice of sign and respective integration contours that pass through the support of delta functions we will get:
\begin{equation}
    B_2(p) \sim \delta(p-2m)\,, \delta(p+2m) \,, \delta(p)
\end{equation}
just as expected. Things become more complicated, when we go to higher dimensions. In $D=3$ the position space, equations are formally solved by:
\begin{equation}
    G(x) = \dfrac{e^{\pm i m x}}{x}
\end{equation}
at least outside $x=0$. Therefore, the three 2-banana functions are:
\be
D=3:  \ \ \ \  \frac{e^{\pm 2imx}}{x^2} \ \ \ {\rm  and} \ \ \ \frac{1}{x^2}
\label{D3solsinx}
\ee
At the same time, the momentum space equation for $B_2$ at generic $D$ is:
\begin{equation}
	\left( 2p^2(p^2-4m^2)\dfrac{d }{dp^2}+(p^2(4-D)-4m^2) \right)B_2(p^2)=0
	\label{mom2}
\end{equation}
and for $D=3$:
\begin{equation}\label{eq:momspaceD3oneloop}
     (p^2-4m^2)\left(2p^2\frac{\p}{\p p^2} + 1\right)B_2(p^2)= (p^2-4m^2)\left(p\frac{\p}{\p p} + 1\right)B_2(p)=0
     \end{equation}
and its naive zero mode is
\begin{equation}
	B_2(p)=\dfrac{1}{p}
\end{equation}
It is in full agreement with the momentum-space integral:
\begin{equation}
	\int d^3q \delta(q^2-m^2)\delta((p-q)^2-m^2)
	\sim \int\int  q^{2}dq \sin\theta d\theta
	\delta(q^2-m^2)\delta(p^2+2pq\cos \theta)
	\sim \dfrac{1}{p}
	\nn
\end{equation}
and is the Fourier transform of the mass independent solution (after proper regularization):
\begin{equation}
    \dfrac{1}{p} \sim \int \dfrac{d^{3}x}{x^2}
\end{equation}
The other two position space functions depend on the mass, hence surely cannot be obtained that way. Instead, to find these, we should also recall that any solution to:
\begin{equation}
    \left(p\frac{\p}{\p p} + 1\right)B_2(p) = \delta(p^2-4m^2)
\end{equation}
will be a solution to \eqref{eq:momspaceD3oneloop}. This produces another two solutions: 

\begin{equation}
B_2(p)=\dfrac{\theta(p-2m)}{p} \, , \dfrac{\theta(p+2m)}{p}
\end{equation}
and these are exactly the Fourier transforms of the two remaining coordinate-space solutions. In the inverse Fourier transform one must take care of the integration contour, since for real momenta $p$ is always positive and hence $\theta(p+2m)$ won't make much sense. Still, one can integrate over contours with complex momenta and recover the proper integrals. We hope to deal with these issues properly in the future, as this seems to be crucial in understanding the problem of factorization. 
\\

The point of this example is to demonstrate that in order to properly compare the solutions in position and momentum space, we should take into account generalized solutions, like delta-functions \cite{10.1007/BFb0068144} and contour choices in the inverse transform.


\section{Beyond banana
\label{beyond}}

In this section we speculate on the possible extension of our method to generic Feynman diagrams and provide the most basic example of the triangle graph. As anticipated in the introduction one can actually consider diagrams without integration over internal vertices. In other words all vertices are external. This means that in position space the corresponding functions is just a product of propagators. It depends on $V$  positions of vertices, however, since propagators are Lorenz invariant, it only depends on $V(V-1)\over 2$ invariant. However in momentum space it is still an $L$-loop integral and depends on $V-1$ external momenta, inflowing into vertices, or again $V(V-1)/2$ invariant $t_{i,j}=(p_i \cdot p_j)$ constrained by $\sum_i p_i=0$. Integration over a coordinate of a single vertex will set the corresponding momentum to zero. Thus, this momentum functions are actually more general, it includes all integrals with given graph topology as special values. On the other hand each of the propagators satisfies the equations of motion. If, instead of propagators, we have multiple lines (bananized propagators) it satisfies the equations given in this paper instead. We will not go further into the development of this approach in this work. Instead, we consider an example. 
\\

Consider the \textbf{Massless triangle diagram}. In position space it is given just by:
\begin{equation}
	T(x_1,x_2,x_3):= G(x_1-x_2)G(x_2-x_3)G(x_1-x_3)
\end{equation}
We again consider here, that $G$ satisfy homogeneous equations, i.e. are cut propagators. The function actually depends on Poincare invariants:
\begin{equation}
	x_{ij}= \sqrt{(x_i-x_j)^\mu (x_i-x_j)_\mu}
\end{equation}
Hence we may write:
\begin{equation}
	T(x_1,x_2,x_3)=T(x_{12},x_{23},x_{13})
\end{equation}
Meanwhile, in momentum space we have:
\begin{equation}\label{eq:TriangleFourier}
	\begin{split}
			T(x_{12},x_{23},x_{13}) &= 	\int e^{i x_1^\mu p_1^\mu + i x_2^\mu p_2^\mu + i x_3^\mu p_3^\mu} \delta \left(p_1+p_2+p_3 \right) \tilde{T}(	t_1=p_1^2 , t_2=p_2^2 , t_3=p_3^2) = 
			\\ &=	\int e^{i  k_1^\mu (x_1^\mu - x_2^\mu)+ i k_2^\mu (x_2^\mu-x_3^\mu)} \tilde{T}(	t_1=k_1^2 , t_2=k_2^2 , t_3=(k_1+k_2)^2)
	\end{split}
\end{equation}
where the momentum loop integral is:
\begin{equation}
    \tilde{T}(	t_1,t_2,t_3)= \int \dfrac{d^Dq}{q^2(q-k_1)^2)(q-k_2)^2}
\end{equation}
It can be rewritten as a parametric integral:
\begin{equation}
	\tilde{T}(	t_1=p_1^2 , t_2=p_2^2 , t_3=p_3^2) = \oint d\alpha_1 d\alpha_2 d \alpha_3 \dfrac{U^{3-D}}{F^{3-D/2}}
\end{equation}
with
\begin{equation}
	\begin{split}
		F&= \alpha_2\alpha_3 t_1 + \alpha_1\alpha_3 t_2 + \alpha_1\alpha_2 t_3\\
		U&=\alpha_1+\alpha_2+\alpha_3
	\end{split}
\end{equation}
This integral satisfies Picard-Fuchs equations, which are now three equations of order one:
\begin{equation}\label{eq:trianglePF}
	\begin{split}
		&\Big\{	-2 t_1\left(t_1^2+t_2^2+t_3^2-2 t_2 t_1-2 t_3 t_1-2 t_2 t_3\right) \dfrac{\partial}{\partial t_1} + 
		\\
		& \hspace{3cm}+ \Big((-4 + D) (t_2 - t_3)^2-(-2 + D) t_1^2 + 2 t_1 (t_2 + t_3)\Big) \Big\} \tilde{T}(t_1,t_2,t_3) = 0 \\
		&\Big\{	-2 t_2\left(t_1^2+t_2^2+t_3^2-2 t_2 t_1-2 t_3 t_1 -2 t_2 t_3\right) \dfrac{\partial}{\partial t_2}  + 
		\\
		& \hspace{3cm}+  \Big((-4 + D) (t_1 - t_3)^2 - (-2 + D) t_2^2 + 2 t_2( t_3 + t_1)\Big)\Big\} \tilde{T}(t_1,t_2,t_3) = 0
		\\
		&\Big\{	-2 t_3\left(t_1^2+t_2^2+t_3^2-2 t_2 t_1-2 t_3 t_1 -2 t_2 t_3\right) \dfrac{\partial}{\partial t_3}  + 
		\\
		&\hspace{3cm}+ \Big((-4 + D) (t_1 - t_2)^2 - (-2 + D) t_3^2 + 2 t_3( t_2 + t_1)\Big) \Big\} \tilde{T}(t_1,t_2,t_3) = 0
	\end{split}
\end{equation}
One the other hand the position space function satisfies:
\begin{equation}
	\left\{ \begin{split}
		&\Box_{x_{12}} T(x_1,x_2,x_3)=  0
		\\
		&\Box_{x_{23}} T(x_1,x_2,x_3)=  0
		\\
		&\Box_{x_{13}} T(x_1,x_2,x_3)=  0
	\end{split}\right.
\end{equation}
To complicated part is to make the Fourier transform these operators, since they are written in terms of $x_{ij}$. We need to transform:
\begin{equation}
	\left( \dfrac{\partial^2}{\partial x_{12}^2} + \dfrac{D-1}{x_{12}}\dfrac{\partial}{\partial x_{12}} \right) T(x_1,x_2,x_3) =0
\end{equation}
which we can rewrite as:
\begin{equation}
	\dfrac{1}{x_{12}^2}\left(\Lambda_{12}^2+ ({D-2})\Lambda_{12}  \right)T(x_1,x_2,x_3) =0
\end{equation}
to momentum space. Here we have denoted:
\begin{equation}
	\Lambda_{ij} = x_{ij} \dfrac{\partial}{\partial  x_{ij}}
\end{equation} the technical difficulty that we face here is that the momenta are coupled to $x_{i}^\mu$ but not to the $x_{12}^2$. There might be several ways to resolve this issue, here we just take the naive one. Consider the following relations:
\begin{equation}\label{eq:tomodules}
\begin{split}
	&(x_{1} -x_2)^\mu \left( \dfrac{\partial}{\partial x_{1}^\mu}+\dfrac{\partial}{\partial x_{2}^\mu}\right) = -\frac{ \left(x_{23}^2-x_{12}^2-x_{13}^2\right)}{2 x_{13}^2} \Lambda_{12}+\frac{
		\left(x_{13}^2-x_{12}^2-x_{23}^2\right)}{2 x_{23}^2} \Lambda_{23}
	\\
	&(x_{2} -x_3)^\mu \left( \dfrac{\partial}{\partial x_{2}^\mu}+\dfrac{\partial}{\partial x_{3}^\mu}\right) = -\frac{ \left(x_{23}^2-x_{12}^2-x_{13}^2\right)}{2 x_{13}^2}\Lambda_{13}+\frac{
		\left(x_{13}^2-x_{12}^2-x_{23}^2\right)}{2 x_{23}^2}\Lambda_{23}
	\\
	&x_1^\mu  \dfrac{\partial}{\partial x_1^\mu} + x_2^\mu  \dfrac{\partial}{\partial x_2^\mu}+x_3^\mu  \dfrac{\partial}{\partial x_3^\mu} = \Lambda_{12}+\Lambda_{23}+\Lambda_{13}  
\end{split}
\end{equation}
This is system is non-degenerate, so one can express $\Lambda_{12},\Lambda_{23},\Lambda_{13}$ from it. On the other hand, the operators on the l.h.s. can be now transformed to momentum space. In particular:
\begin{equation}
	\begin{split}
		(x_{1} -x_2)^\mu \left( \dfrac{\partial}{\partial x_{1}^\mu}+\dfrac{\partial}{\partial x_{2}^\mu}\right) &\quad \rightarrow \quad \dfrac{\partial}{\partial k_1^\mu} k_1^\mu \phantom{+\dfrac{\partial}{\partial k_1^\mu} k_1^\mu} \quad \rightarrow \quad -\left(D+t_1 \dfrac{\partial}{\partial t_1} +(t_2-t_1-t_3) \dfrac{\partial}{\partial t_3} \right)
		\\
		(x_{2} -x_3)^\mu \left( \dfrac{\partial}{\partial x_{2}^\mu}+\dfrac{\partial}{\partial x_{3}^\mu}\right)  &\quad \rightarrow \quad \dfrac{\partial}{\partial k_2^\mu} k_2^\mu \phantom{+\dfrac{\partial}{\partial k_1^\mu} k_1^\mu}
		\quad \rightarrow \quad  -\left(D+t_2 \dfrac{\partial}{\partial t_2} +(t_1-t_2-t_3) \dfrac{\partial}{\partial t_3} \right)
		\\
		\hspace{-1cm }x_1^\mu  \dfrac{\partial}{\partial x_1^\mu} + x_2^\mu  \dfrac{\partial}{\partial x_2^\mu}+x_3^\mu  \dfrac{\partial}{\partial x_3^\mu}x_1^\mu&\quad \rightarrow \quad  \dfrac{\partial}{\partial k_1^\mu} k_1^\mu+\dfrac{\partial}{\partial k_2^\mu} k_2^\mu \quad  \hspace{-0.45em}\rightarrow \quad  -\left(2D+ t_1 \dfrac{\partial}{\partial t_1} +t_2 \dfrac{\partial}{\partial t_2}-2t_3 \dfrac{\partial}{\partial t_3}\right)
	\end{split}
\end{equation}
and 
\begin{equation}
	\begin{split}
			&x_{12}^2 \rightarrow - \dfrac{\partial^2}{\partial k_1^\mu k_1^\mu}
			\\
			&x_{23}^2  \rightarrow -   \dfrac{\partial^2}{\partial k_2^\mu k_2^\mu}
				\\
			&x_{13}^2  \rightarrow -   \left( \dfrac{\partial }{\partial k_1^\mu} + \dfrac{\partial }{\partial k_2^\mu} \right)^2
	\end{split}
\end{equation}
which also can be rewritten in terms of invariants $t_1,t_2,t_3$. Therefore, we can write a Fourier transform for the whole operator:
\begin{equation}
	\Lambda_{ij}^2+ ({D-2})\Lambda_{ij}  \quad  \longrightarrow \quad \hat{E}_{ij}(t_1,t_2,t_3) 
\end{equation}
The goal of this section is to demonstrate that in principle everything works for other diagrams as well. The constructed operators, indeed, annihilate the triangle integral. However, we see that we are again faced with a problem of factorization, a more involved one this time. In particular, after inverting the relation \eqref{eq:tomodules} we get large denominators, which, when brought to a common denominator, lead to equations of degree $12$ in $x$. Clearly after Fourier transform this is way too large compared to the Picard-Fuchs equation \eqref{eq:trianglePF}. That being said, this is most likely due to taking a very naive approach to Fourier transform, which takes the $\delta$-function into account straight away. Instead, one should perhaps use the $p_i^\mu$ variables in \eqref{eq:TriangleFourier} and then carefully pull them through the $\delta$-function. We leave this subject for the future.
\\

The summary of this section is that {\bf{the position space approach works for other diagrams}} as well. However, further work is required to adapt it to the technical issues that arise in these cases.

\section{Conclusion}\label{sec:conclusion}

Our main task in this paper was to formulate the factorization problem (\ref{factoriz}),
relating the two obviously existing equations on the Feynman diagrams. One is in coordinate space, where the Feynman diagram is just a product of Green functions,
and the other in momentum space, where it is an integral over Feynman $\alpha$-parameters,
handled be the standard methods of algebraic geometry.
Associated with these equations are two differential operators $\hat{\mathcal{E}}_x^{(n)}$ and $\widehat{PF}$ --
which are themselves not very easy to find.
As the Picard-Fuchs $\widehat{PF}$ is quite well studied, our main focus is the position space operator.
We explain that it is a part of an interesting problem (\ref{factoriz}),
which is at least as rich as the PF story --
and illustrate this by the simplest example of banana diagrams.\\

We have explained the rather interesting factorization patterns that happen between the Fourier transform $\hat{E}_t$ of the position space operators  and the $\widehat{PF}$ operator. As usual the complexity of the problem depends on whether the masses are generic or all equal. The equal mass case, which was already extensively studied in \cite{Mishnyakov:2023wpd} serves as a nice demonstration of the validity and advantage of our approach. One of its main features is that it is straightforward and based on first principles - the equation of motions. We hope that this approach will attract more attention in the future.
\\

There are certain aspects of the problems that we have not touched in this paper. First are peculiarities related to causality and differences between Minkowski and Euclidean signatures, and possible divergences of momentum space integrals. It would be important to furnish our consideration to include these aspects as well. Further, we considered only the cut propagators, hence, homogeneous differential equations, and did not discuss the restoration of the r.h.s in the position space approach. With the homogeneous equations, we do not consider the full space of solutions.
\\

Still, even with this in mind, there are many questions left, about the equation and their Fourier relation to Picard-Fuchs equations, and factorization. Among those are:
\begin{itemize}
	\item The possibility of the derivation of the general structure of the differential operator from the recursive or determinant formula of sec. \ref{sec:coord}. As we have seen the vanishing mass operator can be fully derived, while for the generic mass nothing is known so far. The equal mass has the middle ground, one can derive its structure order by order, however, for subleading terms it quickly becomes complicated. This would also allow understanding the degrees of these operators.
	\item Determining the order of the Picard-Fuchs operator directly from the position space operator. For now, we don't understand how to determine, which part of  $\hat{E}^{(n)}_t$ factors from the right, without referring to the Picard-Fuchs equations.
	\item Understanding the respective space of solutions. The discrepancy in the orders of the differential operators in momentum space raises the question about mapping the respective solutions to position space and comparing those. Things get more complicated since we need to consider solutions which are distributions. We briefly touched upon this subject in sec. \ref{sec:factorization}. However, clearly a more systematic and careful study of this subject is needed. 
	\item Generalizing to other diagrams. We presented here a potential generalization of the position space approach to the triangle graph. We hope to expand on this in future works.
	\item The possibility of generalization to other equations of motion. In particular, we are certainly not bound to consider equations of second order. We could generalize our approach to equations such as $\Box^2+m^2$ and so on. Furthermore, coordinate space is well suited and even somewhat obligatory for considering curved space, therefore a possibility of an extension to $dS$ or $AdS$ correlators is among the immediate further developments \cite{Chowdhury:2023arc,Cacciatori:2024zrv,Cacciatori:2024zbe}. 	 
\end{itemize}

\section*{Acknowledgements}
V.M. is grateful to Anna-Laura Sattelberger for valuable discussions.
The work of A.M., M.R. and P.S. is supported by the Russian Science Foundation (Grant No.20-71-10073). Nordita is supported in part by NordForsk.

\appendix

\section{More examples of differential equations}\label{Appendix}
\subsection{Equal mass Picard-Fuchs operators}
\label{AppendixA1}
\paragraph{6 loops}
\begin{equation}
	\begin{aligned}
		\Big\{&-64 t^3 (-49 m^2 + t) (-25 m^2 + t) (-9 m^2 + t) (-m^2 + t)\frac{\partial^6}{\partial t^6}+96 t^2 \Big(-11025 (4 + D) m^8 - 12916 (-10 + D) m^6 t + \\
		&+1974 (-16 + 3 D) m^4 t^2 + 84 (22 - 5 D) m^2 t^3 + 7 (-4 + D) t^4\Big)\frac{\partial^5}{\partial t^5}-16 t \Big(33075 (2 + D) (4 + D) m^8 + \\
		&+4 (-740052 + D (98761 + 11636 D)) m^6 t +
		6 (234504 + D (-88952 + 7433 D)) m^4 t^2 - \\
		&-84 (1476 + D (-703 + 82 D)) m^2 t^3 + 35 (-4 + D) (-18 + 5 D) t^4\Big)\frac{\partial^4}{\partial t^4}+8 \Big(-11025 D (2 + D) (4 + D) m^8 + \\
		&+4 (2 + D) (627648 + D (-242728 + 18705 D)) m^6 t +
		18 (-405280 + D (225752 + D (-33266 + 589 D))) m^4 t^2 - \\
		&-84 (-18 + 5 D) (752 + D (-356 + 39 D)) m^2 t^3 +
		35 (-4 + D) (-10 + 3 D) (-24 + 7 D) t^4\Big)\frac{\partial^3}{\partial t^3}+\\
		&+16 (-3 + D) \Big(5 D (2 + D) (-13172 + 3485 D) m^6 +
		3 (284704 + D (-79460 + D (-17299 + 4520 D))) m^4 t + \\
		&+3 (-113472 + D (81808 + D (-18380 + 1219 D))) m^2 t^2 -
		7 (-4 + D) (600 + D (-373 + 58 D)) t^3\Big)\frac{\partial^2}{\partial t^2}-\\
		&-24 (-3 + D) (-8 + 3 D) \Big(D (10852 - 5505 D + 575 D^2) m^4 -
		2 (-14 + 5 D) (-540 + D (161 + D)) m^2 t - \\
		&-7 (-4 + D) (-3 + D) (-20 + 7 D) t^2\Big)\frac{\partial}{\partial t}-\\
		&-8 (-3 + D) (-5 + 2 D) (-8 + 3 D) (-7 + 3 D) (-12 +
		5 D) ((4 + 5 D) m^2 + (-4 + D) t)\Big\}B_7=0
	\end{aligned}
	\label{eq_6_loops_pf}
\end{equation}

\paragraph{7 loops}
\begin{equation}
	\begin{aligned}
		\Big\{&  128  (t - 64 m^{2})  (t - 36 m^{2})  (t - 16 m^{2})  (t - 4 m^{2})  t^{4}\frac{\p^7}{\p t^7}-1792  t^{3}  \Big(\left(D - 4\right) t^{4} + 30 (13 - 3 D) m^2 t^{3} +\\
		&+ \left(2184 (-5 + D) m^4\right) t^{2} - 13120 (-7 + D) m^6 t - 147456 m^{8}\Big)\frac{\p^6}{\p t^6}+\\
		&+ 64  t^{2}  \Big(7 (-4 + D) (-84 + 23 D) t^{4} -168 (1068 + D (-514 + 61 D)) m^2 t^{3} +\\
		&+120 (30357 + D (-12448 + 1201 D)) m^4 t^{2} + -256 (75396 + D (-20168 + 737 D)) m^6 t -\\
		&-442368 (-8 + D) (4 + D) m^8\Big)\frac{\p^5}{\p t^5} - 64  t  \Big(35 (-4 + D) (-7 + 2 D) (-24 + 7 D)t^{4} +\\
		&+ -420 (-11 + 3 D) (288 + D (-139 + 16 D)) m^2 t^{3} +\\
		&+ 240 (-74469 + D (46620 + D (-8951 + 491 D))) m^4 t^{2} +\\
		&+ 384 (128764 + D (-42794 + D (-581 + 711 D))) m^6 t +\\
		&+ 147456 (2 + D) (4 + D) (-9 + 2 D) m^8\Big)\frac{\p^4}{\p t^4} +\\
		&+ 8 \Big(7 (-4 + D) (-33600 + D (30896 + D (-9468 + 967 D))) t^{4} -\\
		&-120 (302976 + D (-320392 + D (124586 + D (-20984 + 1279 D)))) m^2 t^{3} -\\
		&-48 (-5725512 + D (4748440 + D (-1281382 + 3 D (35752 + 441 D)))) m^4 t^{2} +\\
		&+768 (2 + D) (-167728 + 3 D (36048 + D (-6742 + 295 D))) m^6 t-\\
		&-147456 D (2 + D) (4 + D) (-10 + 3 D) m^8\Big)\frac{\p^3}{\p t^3} -\\
		&-112 \Big((-4 + D) (-3 + D) (-20 + 7 D) (672 + D (-424 + 67 D)) t^{3} -\\
		&-90 (95808 + D (-101240 + D (38318 - 6000 D + 309 D^2))) m^2 t^{2} -\\
		&-24 (-1055360 + D (607836 + D (1502 + 9 D (-5278 + 691 D)))) m^4 t -\\
		&-192 D (2 + D) (14672 + 9 D (-974 + 141 D)) m^6\Big)\frac{\p^2}{\p t^2}+\\
		&+8 \Big((-3 + D) (-8 + 3 D)  (-4 + D) (-23520 + D (25364 + D (-9107 + 1089 D))) t^{2} +\\
		&+ 4 (-235200 + D (239936 + D (-78412 + D (7199 + 327 D)))) m^2 t -\\
		&12 D (66964 + D (-58420 + 3 (5177 - 381 D) D)) m^4\Big)\frac{\p}{\p t}-\\
		&-8 (-3 + D) (-5 + 2 D) (-8 + 3 D) (-7 + 3 D) (-12 + 5 D) (-16 + 7 D)  (\left(D - 4\right) t + 6 m^{2} D + 4 m^{2})\Big\}B_8=0
	\end{aligned}
	\label{eq_7_loops_pf}
\end{equation}
\subsection{Generic mass Picard-Fuchs operators}
\label{AppendixA2}
\paragraph{Two loops. Generic $D$. Fixed, but different masses.}
For general $D$ and $m_1=1, m_2=3, m_3 =13$ the equation is
\begin{equation}
	\begin{aligned}
		\Big\{
		& 8  (t - 289)  (t - 225)  (t - 121)  (t - 81)  t^{3}  (t^{2} D - 7 t^{2} + 358 t D - 1074 t - 75735 D + 328185)\frac{\p^4}{\p t^4}-\\
		&-4  t^{2}  \Big(5 t^{6} D^{2} - 71 t^{6} D - 358 t^{5} D^{2} + 252 t^{6} + 14678 t^{5} D - 969005 t^{4} D^{2} - 89500 t^{5} + 10045419 t^{4} D +\\
		&+ 244809676 t^{3} D^{2} - 14649704 t^{4} - 3693022716 t^{3} D - 8936441253 t^{2} D^{2} + 9739896840 t^{3} +\\
		&+ 377700774975 t^{2} D - 2060785248390 t D^{2} - 1346616328788 t^{2} - 5289387383610 t D +\\
		&+ 144800024230125 D^{2} + 58271343599700 t - 627466771663875 D\Big)\frac{\p^3}{\p t^3}+\\
		&+2  t  \Big(9 t^{6} D^{3} - 183 t^{6} D^{2} + 1432 t^{5} D^{3} + 1224 t^{6} D - 11098 t^{5} D^{2} - 1450599 t^{4} D^{3} - 2688 t^{6} - 58354 t^{5} D +\\
		&+ 28481835 t^{4} D^{2} + 98793656 t^{3} D^{3} + 422440 t^{5} - 162373902 t^{4} D - 3851159580 t^{3} D^{2} + 14880067659 t^{2} D^{3} +\\
		&+ 261818976 t^{4} + 32140934284 t^{3} D - 60640886385 t^{2} D^{2} - 5529260880 t D^{3} - 71367252240 t^{3} -\\
		&- 1223583331860 t^{2} D + 13117376674710 t D^{2} - 144800024230125 D^{3} + 5005232086752 t^{2} -\\
		&- 47584212495930 t D + 917066820124125 D^{2} - 35712290928600 t - 1254933543327750 D\Big)\frac{\p^2}{\p t^2}-\\
		&- \Big(7 t^{6} D^{4} - 175 t^{6} D^{3} + 2506 t^{5} D^{4} + 1610 t^{6} D^{2} - 48330 t^{5} D^{3} - 708799 t^{4} D^{4} - 6440 t^{6} D + 305732 t^{5} D^{2} +\\
		&+ 20640563 t^{4} D^{3} - 64006804 t^{3} D^{4} + 9408 t^{6} - 678768 t^{5} D - 195227798 t^{4} D^{2} - 88369644 t^{3} D^{3} +\\
		&+ 12875626089 t^{2} D^{4} + 257760 t^{5} + 745412920 t^{4} D + 11590495816 t^{3} D^{2} - 212945640201 t^{2} D^{3} -\\
		&- 224470815030 t D^{4} - 991724736 t^{4} - 74586506928 t^{3} D + 1077217554222 t^{2} D^{2} + 3840578348310 t D^{3} +\\
		&+ 48266674743375 D^{4} + 134825339520 t^{3} - 1256384375496 t^{2} D - 33240883082220 t D^{2} -\\
		&- 498755639014875 D^{3} - 2008076769792 t^{2} + 124339801148640 t D + 1641066941274750 D^{2} -\\
		&- 142849163714400 t - 1673244724437000 D\Big)\frac{\p}{\p t}+\\
		&+(D - 4)  (D - 3)  \Big(t^{5} D^{3} - 21 t^{5} D^{2} + 537 t^{4} D^{3} + 146 t^{5} D - 9129 t^{4} D^{2} - 36898 t^{3} D^{3} - 336 t^{5} +\\
		&+ 46182 t^{4} D + 1658414 t^{3} D^{2} - 27125298 t^{2} D^{3} - 64440 t^{4} - 14325568 t^{3} D + 303183374 t^{2} D^{2} +\\
		&+ 289823841 t D^{3} + 32938832 t^{3} - 566434096 t^{2} D - 20318833881 t D^{2} + 343157026905 D^{3} -\\
		&- 1055840640 t^{2} + 123148237806 t D - 3546877488165 D^{2} - 167339730816 t +\\
		&+ 11673175356810 D - 11904096976200\Big)\Big\}
	\end{aligned}
	\label{pf_2_loops_gen_d_different}
\end{equation}

\bibliographystyle{utphys}
\bibliography{FeynmanRef}{}

\end{document}